\documentclass[acmsmall,screen]{acmart}
\AtBeginDocument{%
  \providecommand\BibTeX{{%
    \normalfont B\kern-0.5em{\scshape i\kern-0.25em b}\kern-0.8em\TeX}}}
\pdfoutput=1 
  
\setcopyright{acmcopyright}
\copyrightyear{20xx}
\acmYear{20xx}

\acmJournal{CSUR}

\renewcommand\footnotetextcopyrightpermission[1]{} 
\setcopyright{none}

\usepackage{chngcntr}

\usepackage{makecell}
\usepackage{subfigure}
\usepackage{supertabular}
\usepackage{tabu}
\usepackage{enumitem}
\usepackage{booktabs}
\usepackage{tabularx}

\usepackage{graphicx}
\usepackage{multirow}

\usepackage{makecell}

\begin{document}

\title{Artificial Intelligence for Complex Network: Potential, Methodology and Application}

\author{Jingtao~Ding$^{*}$, Chang~Liu$^{*}$, Yu~Zheng$^{*}$, Yunke~Zhang$^{*}$, Zihan~Yu, Ruikun~Li, Hongyi~Chen, Jinghua~Piao, Huandong~Wang, Jiazhen~Liu, Yong~Li}
\email{dingjt15@tsinghua.org.cn, liyong07@tsinghua.edu.cn}
\thanks{*These authors contributed equally.}
\affiliation{%
  \institution{Department of Electronic Engineering, Tsinghua University, Beijing National Research Center for Information Science and Technology (BNRist)}
  \country{China}
  }

\renewcommand{\shortauthors}{Ding et al.}

\begin{abstract}

Complex networks pervade various real-world systems, from the natural environment to human societies. The essence of these networks is in their ability to transition and evolve from microscopic disorder--where network topology and node dynamics intertwine--to a macroscopic order characterized by certain collective behaviors. Over the past two decades, complex network science has significantly enhanced our understanding of the statistical mechanics, structures, and dynamics underlying real-world networks. Despite these advancements, there remain considerable challenges in exploring more realistic systems and enhancing practical applications.
The emergence of artificial intelligence (AI) technologies, coupled with the abundance of diverse real-world network data, has heralded a new era in complex network science research. This survey aims to systematically address the potential advantages of AI in overcoming the lingering challenges of complex network research. It endeavors to summarize the pivotal research problems and provide an exhaustive review of the corresponding methodologies and applications. Through this comprehensive survey--the first of its kind on AI for complex networks--we expect to provide valuable insights that will drive further research and advancement in this interdisciplinary field.

\end{abstract}


\maketitle

\section{Introduction}

Complex networks are abstract descriptions of real-world complex systems. For example, cells are described as complex networks of chemicals linked by chemical reactions~\cite{amaral2004complex}; ecological networks link populations together through food chains~\cite{gao2016universal}; and the World Wide Web is a vast virtual network of web pages and hyperlinks~\cite{dorogovtsev2008critical}. These complex networks are just a few of many examples. The local microscopic behavior of these complex networks often shows disorder. However, at the macroscopic scale, they show simple and even symmetrical structures. In order to understand the transition and evolution of complex systems from microscopic disorder to macroscopic order, current complex network studies mainly fall into the following paradigm: the combination of graph theory and statistical mechanics~\cite{albert2002statistical}. They construct the core principle of complex network science, that is, simple random rules and network dynamics together drive the emergence of non-trivial topological structures.

\begin{figure}[t]
    \centering
    \includegraphics[width=0.9\linewidth]{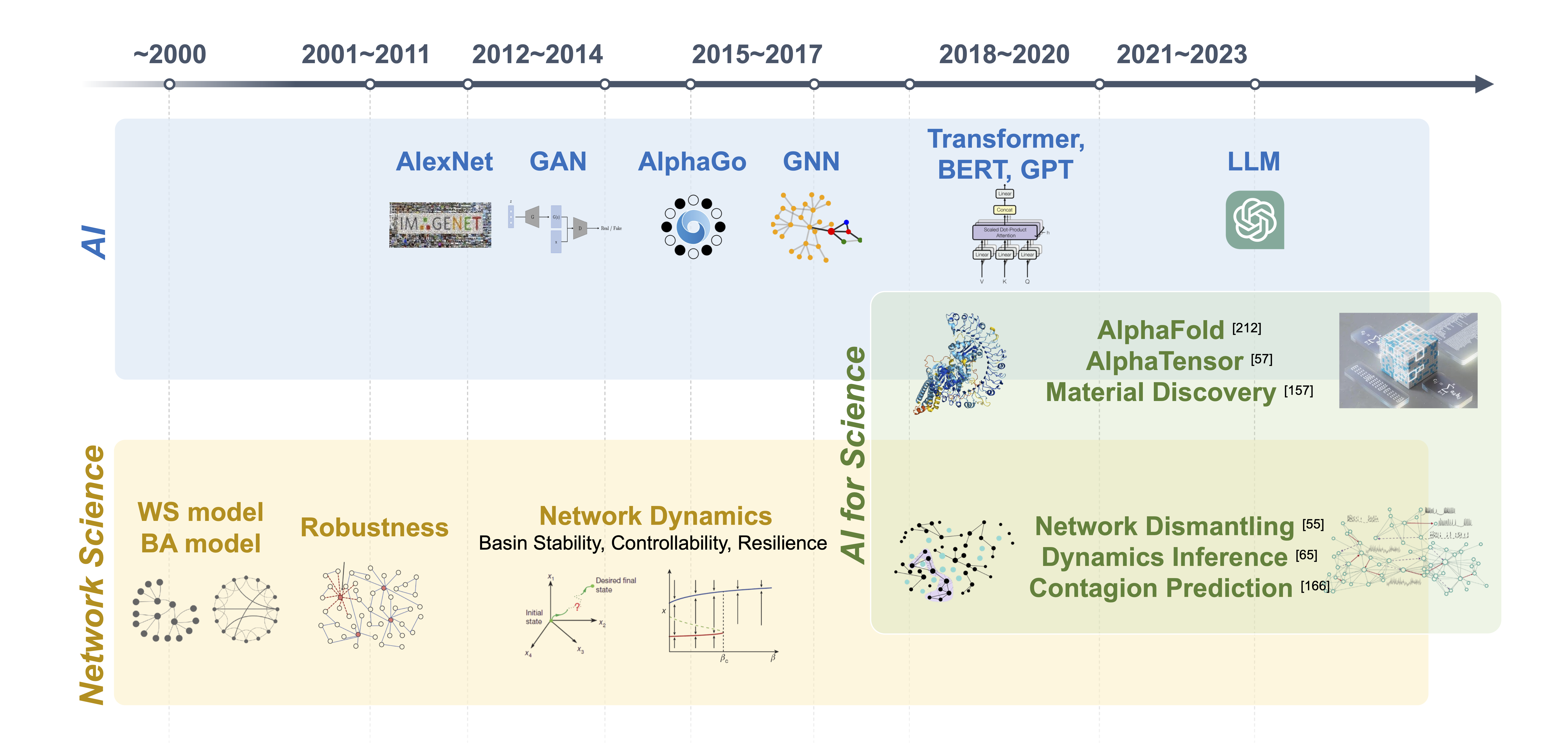}
    \vspace{-2em}
    \caption{Relationship between AI, Complex network and AI for science.}
    \label{fig:intro}
\end{figure}

Early works mainly focused on the topology of the interactions between the components, i.e., the birth-death process of edges on the graph. The two representative works, the Watts-Strogatz (WS) model and the scale-free model~\cite{watts1998collective,barabasi1999emergence}, embody this principle and successfully generate graphs that approach real-world complex networks with high clustering coefficients and small average paths or power-law degree distribution. Despite their success in certain domains~\cite{bianconi2001bose,vazquez2003global,song2005self,song2006origins}, they do not provide a way to model the dynamics of the nodes, i.e., the change in the node's features.  To understand the evolution and non-trivial phenomena of these networks across various fields, a number of models are proposed~\cite{barzel2013universality,harush2017dynamic} and have achieved partial success in understanding some important issues in network science, such as controllability~\cite{liu2011controllability}, resilience~\cite{gao2016universal} and epidemic dynamics~\cite{pastor2015epidemic}. What these models have in common is that they fix the topology of the networks and only consider the interaction between nodes that are linked together. However, this simplification caused discrepancies in real-world networks where topology evolution interplays with the node's dynamics. Recently, we have witnessed an increasing number of models characterizing these so-called coevolving or adaptive dynamics~\cite{santos2021link,liu2023emergence,piao2023human}. These models offered deep insight into how real complex networks run from the mechanistic perspective. However, strong coupled systems are traditionally considered to be intractable in mathematical physics. To obtain analytical solutions, these models adopt reasonable mathematical approximations during calculations. While these approximations help to obtain analytical solutions, they will also lead to discrepancies under certain circumstances. These discrepancies will probably be harmful to some real-world engineering applications. 

On the other hand, the second decade of the 21st century has witnessed revolutionary advancements in the field of artificial intelligence~(AI), marking a new paradigm for scientific discovery. These breakthroughs have unlocked the potential of AI algorithms to accelerate scientific experiments with exceedingly high computational complexity~\cite{zhang2018deep,raissi2019physics,senior2020improved, fawzi2022discovering}. Notably, AlphaFold has revolutionized protein structure prediction~\cite{senior2020improved}, while molecular dynamics simulations have shed light on intricate behaviors of million particles~\cite{zhang2018deep}. AI's prowess extends to solving complex equations like the Navier-Stokes equations~\cite{raissi2019physics}, providing near or even surpassing the accuracy achieved through traditional experiments. Furthermore, AI algorithms have the capacity to unearth novel scientific knowledge and concepts. They have unveiled hidden mathematical and physical phenomena, such as conservation laws~\cite{liu2021machine} and topological invariants~\cite{davies2021advancing}, and have contributed to the discovery of new materials~\cite{goodall2020predicting, kusne2020fly, merchant2023scaling} and proteins~\cite{jumper2021highly}. The most recent success of large language models~(LLMs) has also led to discoveries of improved solve programs for established open mathematical problems~\cite{romera2023mathematical}. This transformative capability of AI represents a paradigm shift in scientific exploration, offering powerful tools for hypothesis generation, data analysis, and identifying emergent scientific principles.  

The rapid development of AI technology provides a new perspective for solving the remaining problems in complex network science~(Fig.~\ref{fig:intro}). First, the network dynamics might be complicated or even unknown in many real-world scenarios. Data-driven and AI methods can help us infer the real-world network dynamic from the real data, without any assumptions. Moreover, coevolving dynamics lead to a strongly coupled and highly nonlinear complex network, infamous for its difficulties in obtaining exact solutions. Deep learning methods are likely to shed light on this problem, due to their unparalleled advantages in solving highly nonlinear dynamic problems. Last but not least, because of the high-dimensional characteristics of complex networks, numerical simulations based on first principles are extremely time-consuming~\cite{zhang2018deep,sanchez2020learning}. AI methods have been proven to be able to significantly reduce computing time in computing tasks such as the spatial layout of complex networks~\cite{sanchez2020learning}. In general, AI methods will provide unique advantages to facilitate complex network research.

There are numerous surveys on complex networks or AI~(Table~\ref{survey}). On the one hand, related surveys on complex networks mainly focus on three categories. These include 1) the comprehensive reviews regarding statistical mechanics~\cite{albert2002statistical}, structure and dynamics~\cite{boccaletti2006complex} of complex networks, the combination with data mining~\cite{zanin2016combining}; 2) the detailed reviews from a theoretical perspective focusing on typical network properties such as coevolving network~\cite{maslennikov2017adaptive}, high-order interactions~\cite{battiston2020networks,battiston2021physics,boccaletti2023structure}, coupled dynamical network~\cite{zou2021quenching}, representation~\cite{comin2020complex,torres2021and}, message passing~\cite{newman2023message}, robustness and resilience~\cite{artime2024robustness}; and 3) the specific reviews on important applications covering coevolution spreading~\cite{wang2019coevolution}, network reconstruction~\cite{squartini2018reconstruction}, vital node itentification~\cite{lu2016vital}, dynamic network recovery~\cite{wu2021recovering}, link prediction~\cite{lu2011link}, structure prediction~\cite{ren2018structure}, control network~\cite{d2023controlling}, dynamics inference~\cite{gao2023data}, signal propagation~\cite{ji2023signal} and hypergraph mining~\cite{lee2024survey}. On the other hand, related surveys on AI focus on specific methodologies including graph embedding~\cite{cai2018comprehensive}, GNN~\cite{wu2020comprehensive}, RL on graphs~\cite{wu2020comprehensive}, generative model~\cite{guo2022systematic} and physics-informed machine learning~\cite{karniadakis2021physics}. With the rising trend of leveraging AI in the scientific research field, several surveysalso discuss this new paradigm in terms of both wide concept of scientific discovery~\cite{xu2021artificial,wang2023scientific} and specific field like ecology~\cite{han2023synergistic}.
However, there is no related survey on the important and promising research field of combining AI with network science. 

In this paper, we systematically discuss the unresolved challenges in complex network research regarding higher-order topological properties, coevolving dynamics, unknown mechanisms and high-dimensional nature. Correspondingly, we give a comprehensive review of the great potential of AI methodologies in tackling the above challenges, including graph embedding, graph neural network, dynamic graph learning, deep generative model, deep reinforcement learning and physics-informed machine learning. To further offer constructive insight regarding the role of AI in complex network science, we summarize six key research problems that focus on the representation, prediction, simulation, inference, generation and control of complex networks, and discuss how various AI methodologies are designed in different problems. Based on our defined taxonomy of research problems, we further illustrate typical applications of AI-enhanced complex network models, covering a wide range of real-world scenarios from natural systems, i.e., ecology networks and biology networks, to human-made systems, i.e., urban networks and social networks. We also discuss several important aspects regarding combining AI with mechanistic approaches, developing complexity theories for AI, and achieving network science discovery in the age of AI.
We expect this first survey on AI for complex networks to provide useful insight for further research and advancement. Our contributions can be summarized as follows:
\begin{itemize}
    \item We comprehensively investigate the complex network research in the age of AI for the first time and examine their core challenges and potentials.
    \item We introduce a novel taxonomy of research problems in complex network science and discuss current development and future directions of corresponding AI methodologies.
    \item We outline the practical applications of AI-enhanced complex network models across various domains and summarize dataset resources that serve as a key component for accelerating data-centric studies in these domains.
\end{itemize}

\section{Background and Challenges}

\begin{figure}[htbp]
\vspace{-1em}
    \centering
    \includegraphics[width=1.0\linewidth]{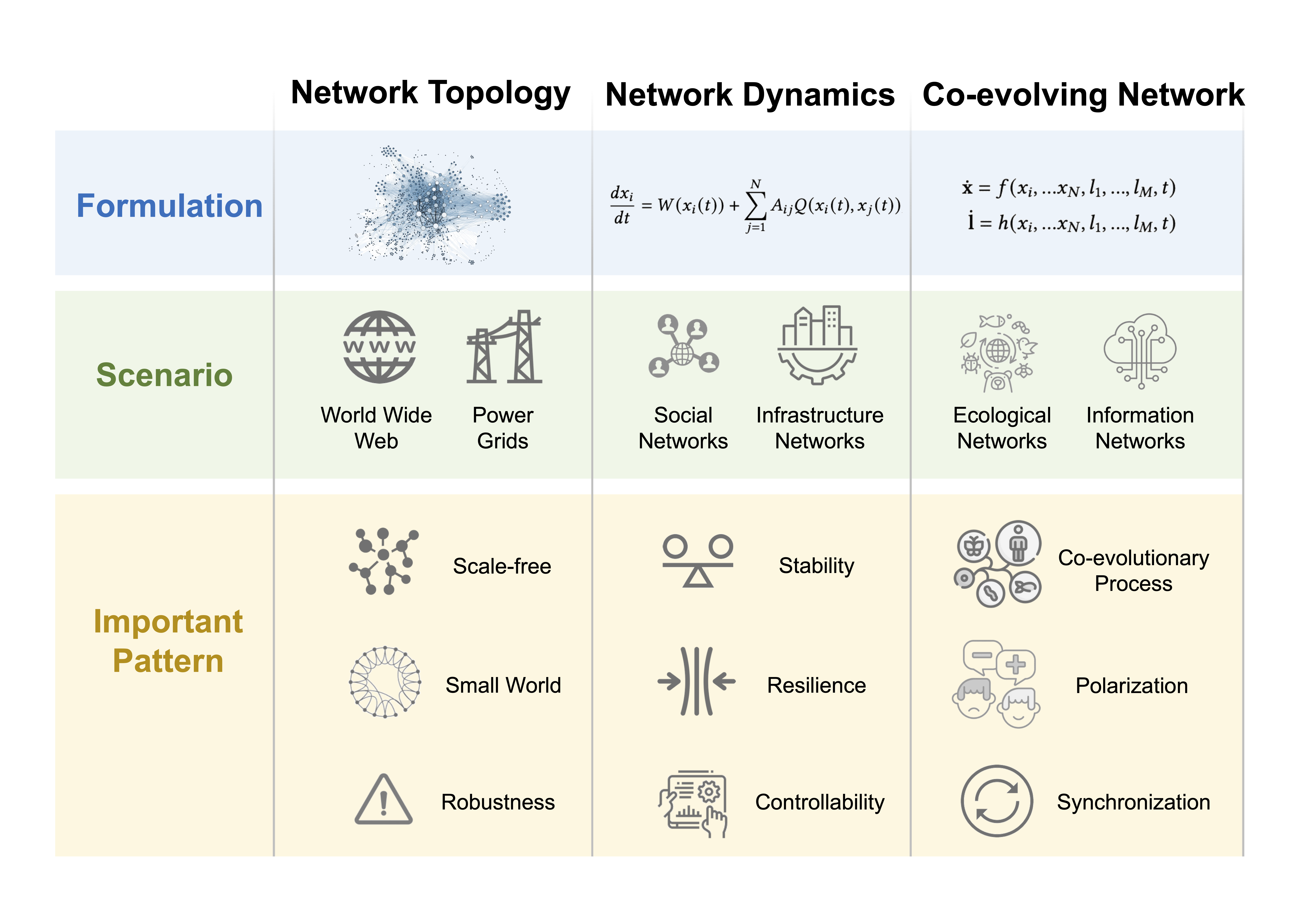}
    \vspace{-3em}
    \caption{Comparison between three mainstreamed types of research on complex networks, focusing on network topology, network dynamics and co-evolving networks, respectively.}
    \label{fig:background}
    \vspace{-1em}
\end{figure}

In this section, we introduce the background and challenges of complex network research. In section~\ref{sec:2.1}, we briefly define key concepts and review three main streams of the literature~(Fig.~\ref{fig:background}). In section~\ref{sec:2.2}, we summarize four major challenges faced by traditional complex network studies.

\subsection{Background} \label{sec:2.1}

\subsubsection{Network topology}

The foundation of complex network studies lies in graph theory. In this context, a \textit{network} (graph) is defined as a tuple $G=(V, E)$ where $V$ is the set of \textit{vertices} (\textit{nodes}), $E$ is the set of \textit{edges} (\textit{links}). Each edge is attached to two vertices $v$ and $w$ (endpoints). Complex network studies primarily focus on networks with nontrivial topological structures, encompassing a wide range of real-world networks such as ecological networks and the World Wide Web. Within this domain, a significant subfield explores the dynamics of edge formation, investigating how edges evolve to give rise to specific, intriguing network topology structures.

Two seminal models, the Watts-Strogatz (WS) network~\cite{watts1998collective} and the Barab{\'a}si-Albert (BA) network~\cite{barabasi1999emergence}, have played pioneering roles in this field. WS model exemplifies the concept of ``small-worldness'', a characteristic found in numerous real-world networks, such as collaboration networks and U.S. power grids. Small-worldness entails a network structure in which the average path length between any two nodes remains relatively short while maintaining a high degree of local clustering akin to random networks. This property is achieved through an initial formation of a regular lattice structure where each node connects to its $k$ nearest neighbors, followed by rewiring each edge to connect to a randomly chosen node with a probability $p$, introducing both randomness and shortcuts to the network. 
The BA model exemplifies the ``scale-free'' property, where the distribution of node degrees (the number of edges that the node has) follows a power-law distribution, signifying that a few nodes possess significantly more connections than the majority. This topology property, widely seen in the Internet and social networks, is achieved through the growth and preferential attachment mechanism, where nodes are incrementally added to the network, and new nodes are more likely to connect to existing nodes with higher degrees. This preferential attachment creates a cumulative advantage for well-connected nodes, resulting in a scale-free network structure. 

Numerous studies have extended the foundational principles of these two models to enhance network growth mechanisms aimed at elucidating the topological characteristics of complex networks. Several models have been devised to adapt and improve upon the WS and BA models to capture better the small-worldness and scale-free properties~\cite{krapivsky2000connectivity, garlaschelli2003universal, papadopoulos2012popularity}. 
Another property that has garnered considerable attention is the robustness of network structures, particularly concerning error and attack tolerance. For instance, in real-world communication networks, local node failures rarely result in losing global network functionality~\cite{claffy1999internet}. This error tolerance has been observed to be a characteristic feature of scale-free networks~\cite{albert2000error} and has subsequently been leveraged in the design of network structures capable of withstanding meticulously designed attack strategies~\cite{dobson2007complex, buldyrev2010catastrophic, watts2002simple}.

\subsubsection{Network dynamics}

The study of complex networks extends its reach beyond static topological properties, delving into the dynamic aspects that govern various networked systems. The history of research on network dynamics finds its roots in the Ising model within statistical physics~\cite{ising1924beitrag}. This model, used to describe the magnetic properties of materials with simple lattice structure, laid the foundation for understanding how local interactions among nodes can give rise to global phenomena within a network. In the context of complex networks, network dynamics investigates how node states evolve under these interactions over time. In a complex network where each node $i$ is characterized by a state $x_i$, the temporal evolution of $x_i$'s will be given by an ordinary differential equation $\frac{d\mathbf{x}}{dt}=F(\mathbf{x},t)$ and the original state $\mathbf{x}(0)=\mathbf{x_0}$. Combined with the network topology, the universal form of network dynamics is expressed as~\cite{barzel2013universality}

\begin{equation}\label{equ:graph_dynamics}
    \frac{dx_i}{dt} = W(x_i(t)) + \sum_{j=1}^NA_{ij}Q(x_i(t),x_j(t)),
\end{equation}
where $W(x_i(t))$ is the self-evolving term, $N$ is the number of nodes, $A_{ij}$ denotes the existence of edge between node $i$ and $j$, and $Q(x_i(t),x_j(t))$ is the interaction term exerted by neighbors of $i$. Under this universal assumption, researchers can delve into specific dynamical properties of complex networks.

Contagion models constitute a significant branch of network dynamics research, with prominent examples being the Susceptible-Infectious-Removed (SIR) models~\cite{harko2014exact} and Susceptible-Exposed-Infectious-Removed (SEIR) models~\cite{li1999global}. 
Understanding the dynamics of contagion processes is critical for predicting the spread of epidemics in cities~\cite{eubank2004modelling}, the diffusion of innovations~\cite{meade2006modelling}, and the flow of information in social networks~\cite{zhao2011social}.
Network dynamics can also explore the resilience of complex networks. For instance, the resilience or collapse of ecological systems in the face of environmental changes or perturbations as well as the sudden market crashes in finance. Specific network dynamics can lead to the spontaneous emergence of phase-transition phenomena~\cite{majdandzic2014spontaneous}. This property can also be predicted~\cite{gao2016universal} and be adopted to control the evolving path of networks~\cite{liu2011controllability}. Network resilience research is pivotal for addressing contemporary challenges to ensure the reliability and stability of crucial infrastructures and services in an increasingly interconnected world~\cite{engsig2024domirank}. 

\subsubsection{Coevolving networks}

The study of complex networks has evolved to encompass the intricate dynamics of co-evolution networks, also known as adaptive networks, a subfield that investigates the intertwined evolution of both network topology and node characteristics. In these networks, the temporal evolution of the network structure itself becomes a vital aspect of analysis, which is aligned with real-world networks, such as the formation and break of social links. A general coevolving network dynamic can be expressed as
\begin{equation}
\begin{split}
   \dot{\mathbf{x}}=f(x_i,...x_N,l_1,...,l_M,t),  \\
   \dot{\mathbf{l}}=h(x_i,...x_N,l_1,...,l_M,t),
\end{split}
\end{equation}
where the node states $x$ evolv with function $f$ and the edge states $\mathbf{l}$ evolve with a generalized function $h$ (\textit{e.g.}, a birth-death process). 

One milestone of coevolving networks studies on ecological networks, which shed light on how species interactions in ecosystems co-evolve with changes in the ecological network structure, such as predator-prey relation in the food web~\cite{drossel2001influence} and mutualistic relation~\cite{allesina2012stability}. This branch of research not only highlights the importance of considering the dynamic nature of species interactions but also demonstrates the profound impact of co-evolutionary processes on ecosystem stability and biodiversity. Another branch of classical studies focuses on information and social networks, particularly in the context of contagious processes, where the dynamics of opinion or disease spread are intertwined with the evolving network topology~\cite{zheleva2009co}. This research illuminates the intricate ways in which the evolution of nodes and network structures drives different phenomena, such as opinion polarization~\cite{vazquez2008generic, liu2023emergence} and the formation of information cocoons in online platforms~\cite{santos2021link, piao2023human}. 
In summary, investigating co-evolution networks represents a crucial dimension of complex network science, emphasizing the simultaneous evolution of both network edges and nodes.

\subsection{Challenges} \label{sec:2.2}

\textit{Characterizing higher-order topological properties in complex networks is challenging.} Presently, most complex network research is centered around networks with pairwise connections. However, interactions in many systems often occur among three or more entities within a network~\cite{battiston2020networks}. To investigate such scenarios, researchers are increasingly exploring ``higher-order'' structures within networks, as they can significantly influence dynamic processes. A key method for explaining higher-order structures in networks involves extending dynamic processes on networks to complex manifolds. Nevertheless, given the involvement of multiple entities, studying higher-order networks inevitably entails the combination of complex manifolds. This introduces a new research perspective on employing algebraic and differential geometric methods to characterize higher-order topology~\cite{battiston2021physics}.

\textit{The dynamic mechanisms within real complex networks remain unknown.} The topological structures and individual-level dynamics in real networks are often highly complex and strongly coupled. The prevailing research paradigms in traditional theoretical models are frequently based on oversimplified assumptions in the form of dynamic functions, rendering them ineffective for establishing comprehensive theoretical frameworks~\cite{barzel2013universality, gao2022autonomous}. Consequently, it becomes challenging to accurately capture numerous significant phenomena in real-world complex networks, including the evolution of diversity in ecological systems~\cite{ives2007stability, calcagno2017diversity}, cascade reactions in multi-layered infrastructure networks~\cite{danziger2022recovery}, resilience analysis in transportation networks~\cite{zhang2015assessing}, and dynamics of infectious diseases in urban networks~\cite{eubank2004modelling}. 
Therefore, enhancing our understanding of the dynamic mechanisms within real-world complex networks and improving predictive accuracy represent crucial challenges in the field of complex network science.

\textit{The coevolution of node states and their interactions is challenging to model theoretically.} This is particularly due to the intricate interplay between topological correlation and the dynamic evolution of node-edge relationships, which is rooted in the strong correlations that exist between the states of nodes and the nature of their interactions. 
Degree correlation refers to the tendency of nodes in a network to connect with others with similar or dissimilar degrees~\cite{park2003origin, vazquez2003growing}. 
The concept of node-edge coevolution highlights the reciprocal relationship between nodes and their connections, adding complexity to modeling~\cite{pastor2001dynamical, krapivsky2000connectivity}. This bidirectional causality requires a framework for node state evolution and network topology reconfiguration. In essence, this challenge lies in capturing the nuanced and often non-linear interdependencies that govern the behavior of complex networks, requiring a multidimensional theoretical model that integrates aspects of topology, node dynamics, and the emergence properties of the network as a whole~\cite{ghavasieh2024diversity}.

\textit{Optimizing high-dimensional complex networks is a formidable challenge.} Optimization problems on high-dimensional complex networks represent one of the most significant challenges in the field of complex networks~\cite{d2023controlling}. Traditional operational research optimization methods involve solving optimization computations on high-dimensional continuous-time differential equations or dynamic programming equations in discrete time, such as high-dimensional Bellman equation systems. However, traditional methods often lack analytical solutions when dealing with high-dimensional complex networks. Additionally, obtaining numerical solutions entails exceedingly high computational complexity and costs. Approximation methods to reduce computational complexity often result in deviations from theoretical optimal solutions. Typical examples of high-dimensional optimization scenarios with extremely high computational complexity include chemical reaction networks, protein networks, brain neural networks, and Ising models.

\section{Potential of AI Technology}
In this section, we give a comprehensive review of related AI technologies that have the potential to tackle the remaining challenges listed in Sec.~\ref{sec:2.2}.

\begin{figure}[htbp]
    \vspace{-1em}
    \centering
    \subfigure[graph embedding]{\includegraphics[width=.49\textwidth]{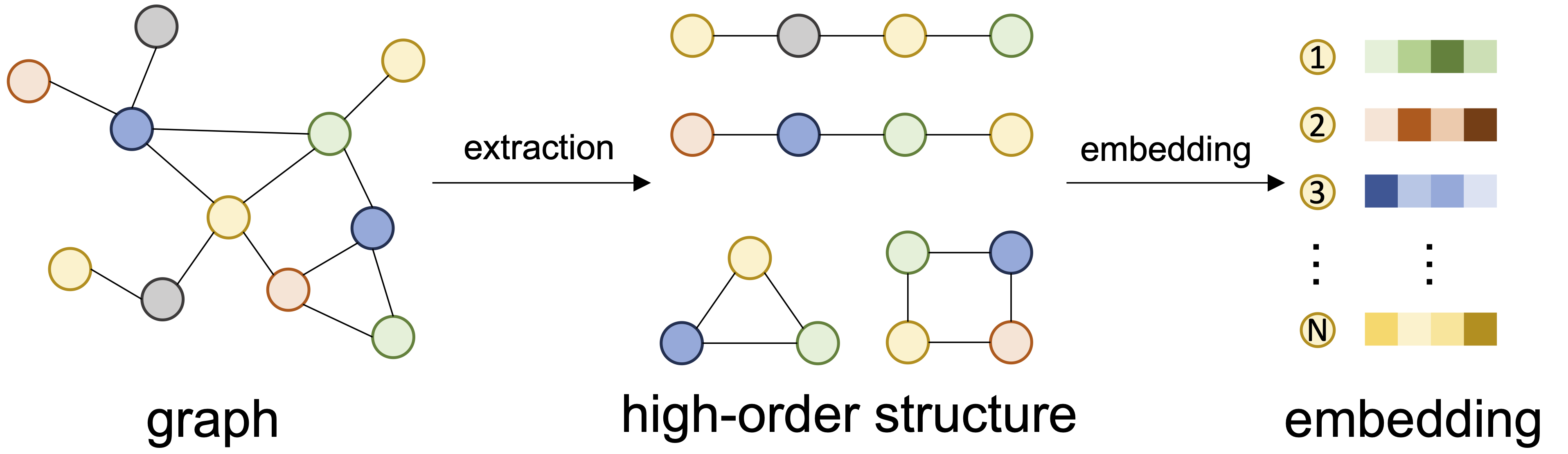}}
    \hfill
    \subfigure[graph neural network]{\includegraphics[width=.49\textwidth]{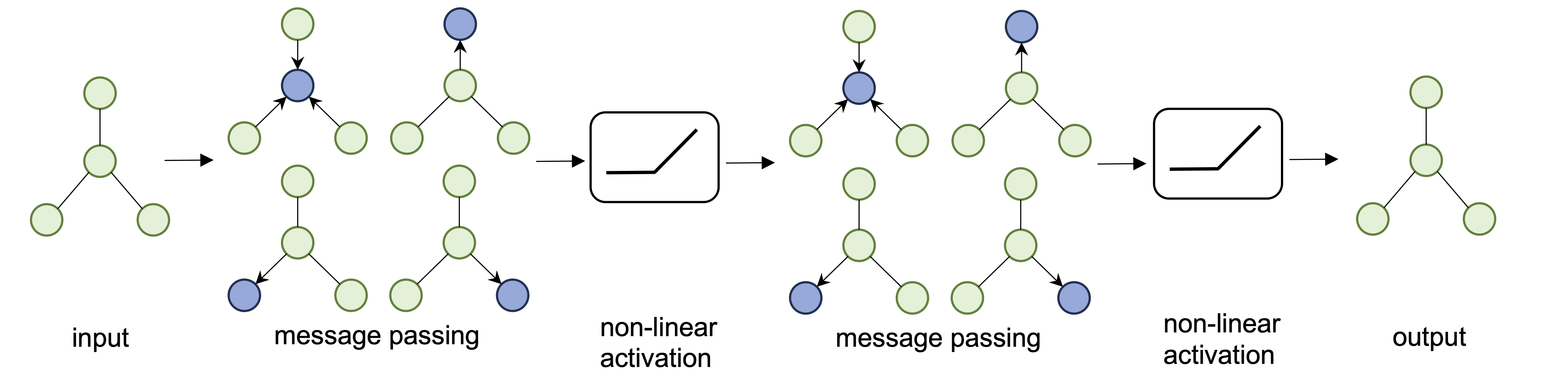}}
    \vfill\vspace{-1em}
    \subfigure[dynamic graph learning]{\includegraphics[width=.46\textwidth]{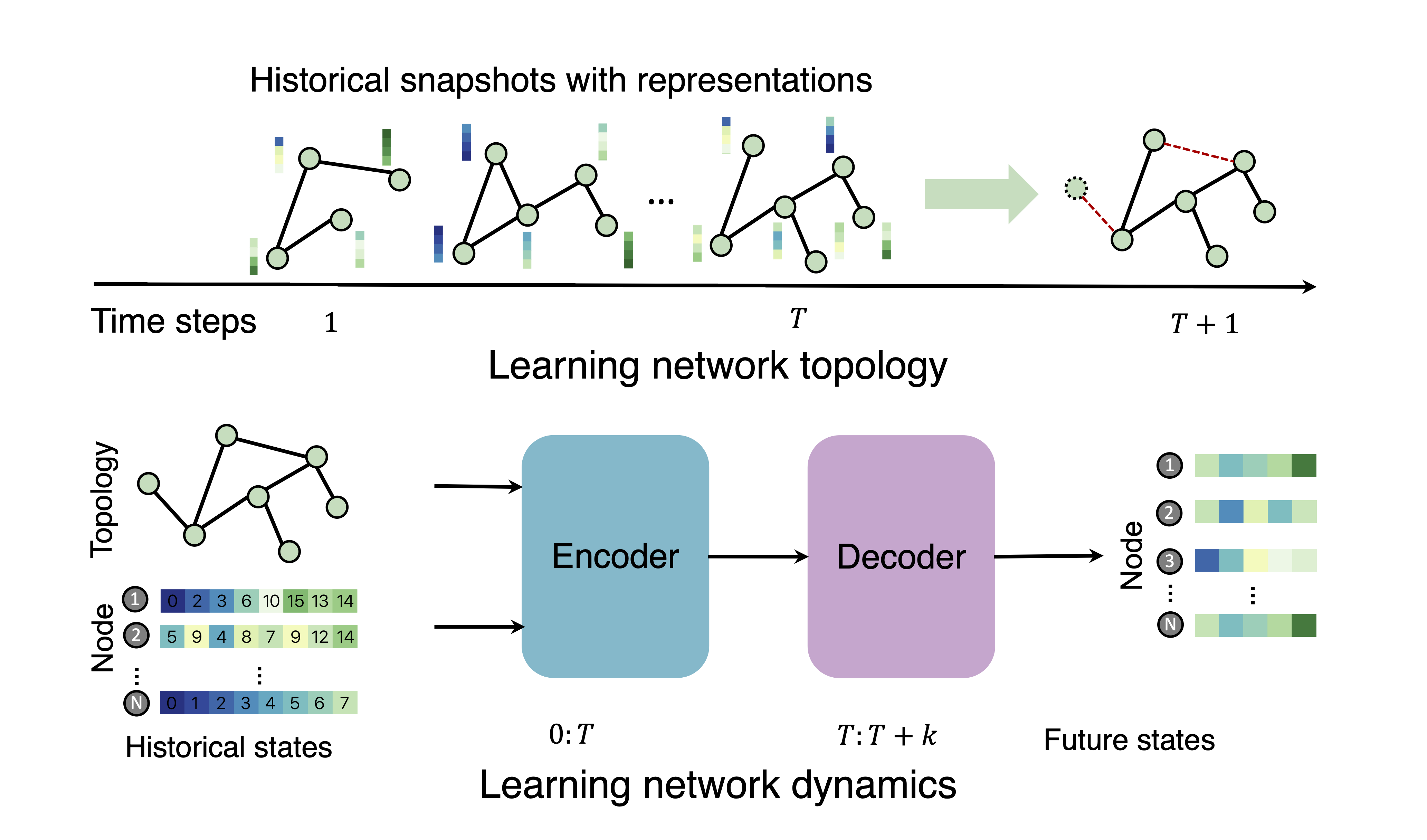}}
    \hfill
    \subfigure[deep generative model]{\includegraphics[width=.5\textwidth]{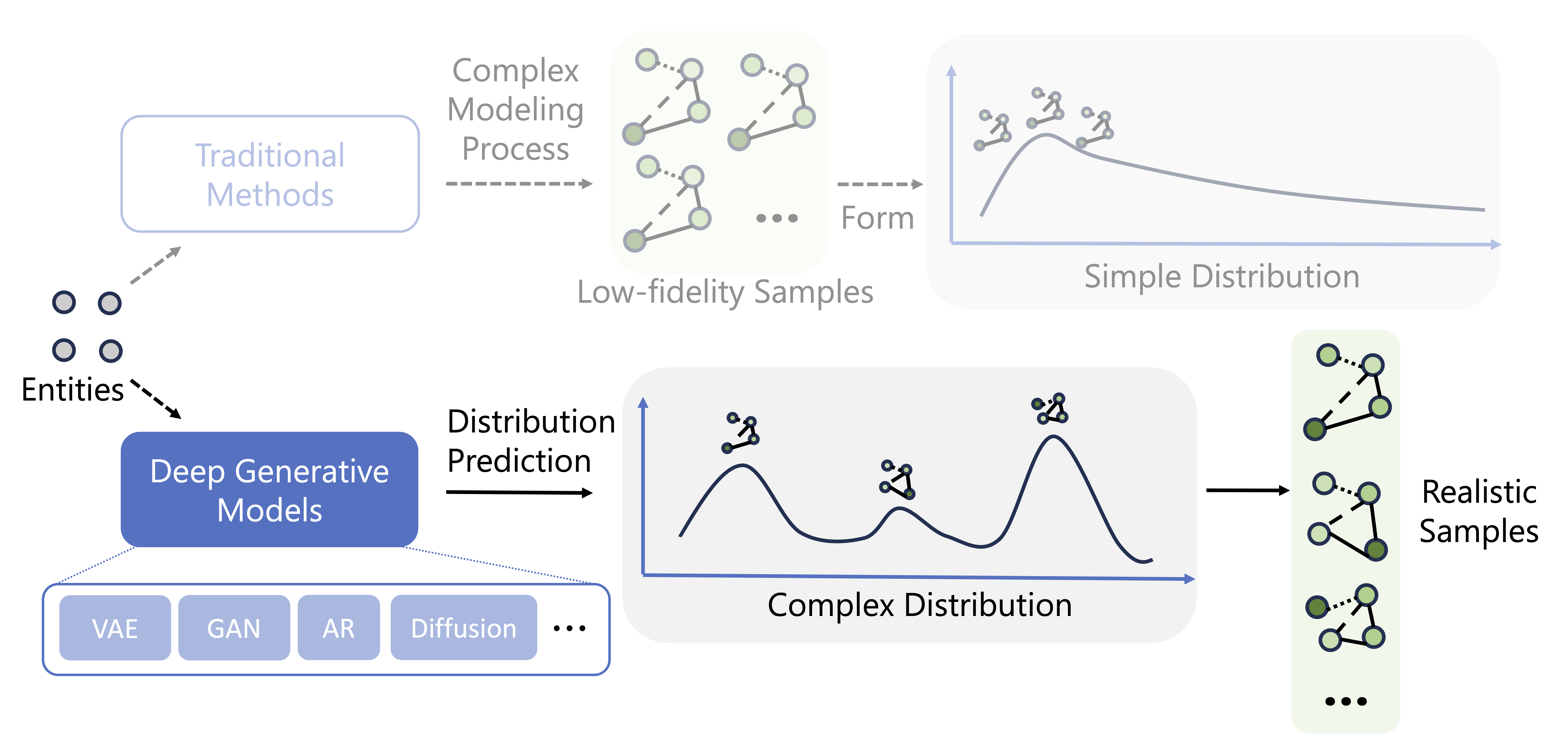}}
    \vfill\vspace{-1em}
    \subfigure[deep reinforcement learning]{\includegraphics[width=.49\textwidth]{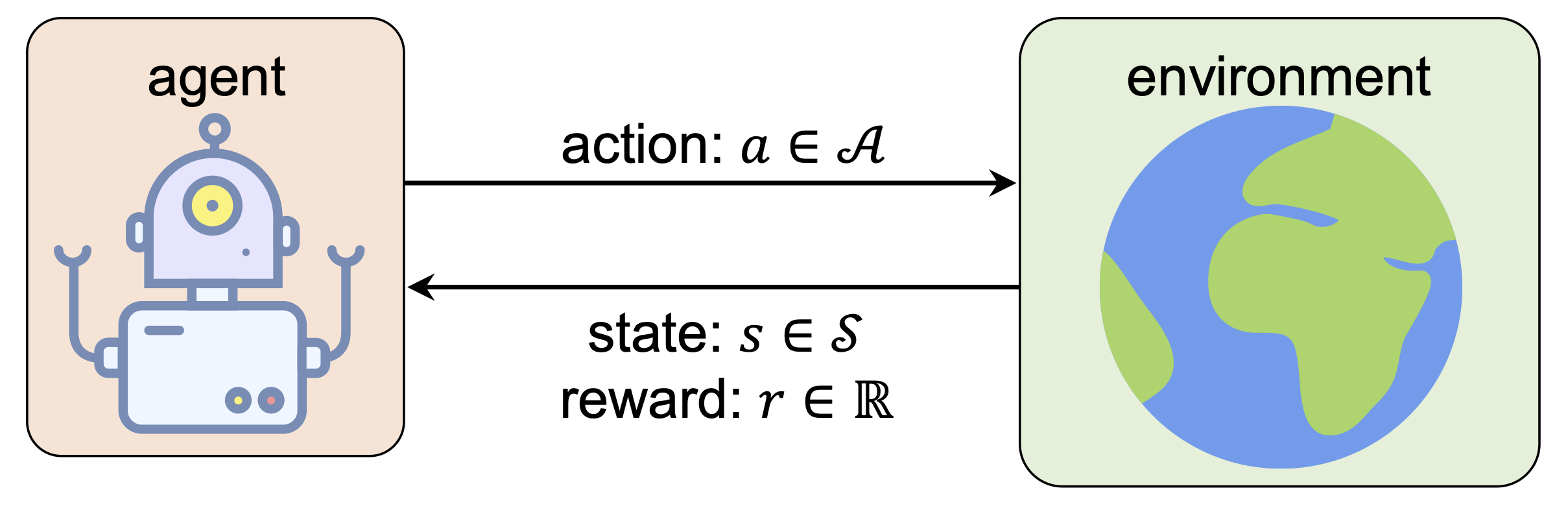}}
    \hfill
    \subfigure[physics-informed machine learning]{\includegraphics[width=.49\textwidth]{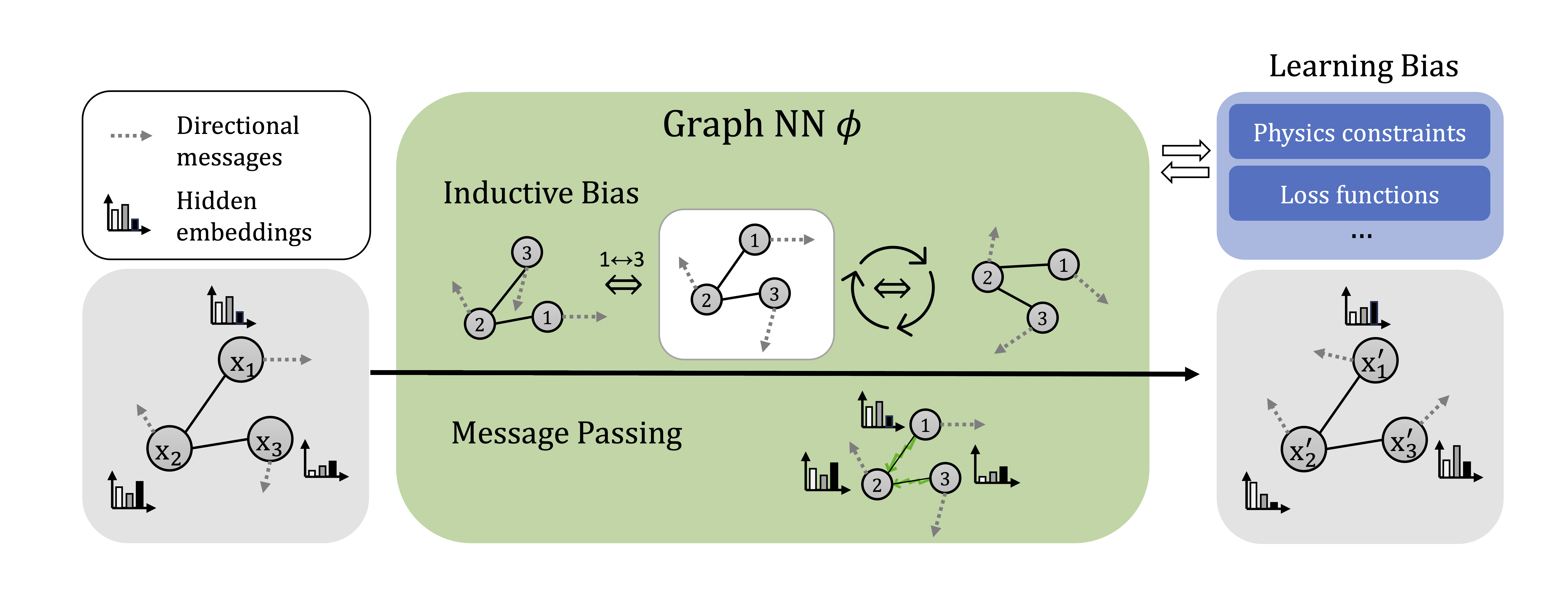}}
    \vspace{-1em}
    \caption{Potential of different AI approaches in tackling unresolved challenges for complex networks. (a) Graph embedding methods first extract high-order structures from graphs, including sampled node sequences and motifs, and then obtain vectorized embeddings for each node, providing a computationally efficient way of capturing high-order topological relationships in complex networks. (b) Graph neural networks~(GNN) use a stack of continuous message-passing and non-linear activation layers to obtain node embeddings, supervised by labeled information, smoothly integrating high-order relationships to complete tasks such as node classification and link prediction. (c) Dynamic graph learning approaches infer future network topology and node states by employing message-passing on the preceding $T$-steps of topology snapshots and leveraging an encoder-decoder architecture to extrapolate the dynamic patterns, respectively, capturing the coevolution of node states and their interactions that are often challenging to model theoretically. (d) Deep generative models directly learn the complex distribution from data and can produce realistic samples by capturing nonlinear relationships and intricate patterns within real networks. (e) Reinforcement learning is a process where an agent receives state and reward from the environment and takes actions to maximize its long-term return by effectively searching the large solution space with functional approximation by deep neural networks, successfully solving the high-dimensional Bellman equation to accomplish complicated decision-making tasks. (f) Physics-informed machine learning in networks integrates information about geometry, structure, and symmetry by devising unique neural message-passing strategies and applying loss functions and constraints guided by certain physics equations, improving accuracy and generalizability.}
    \label{fig:potential}
    \vspace{-1em}
\end{figure}

\subsection{Graph embedding} 

Graph embedding aims to represent nodes or entire graphs in a continuous, low-dimensional vector space, encoding the graph's structure, connectivity, and features into a fixed-size vector for each node or the entire graph.
It is particularly useful in scenarios where nodes and edges of the graph need to be transformed into a format compatible with traditional machine learning algorithms that operate on continuous vector spaces to make it suitable for various downstream tasks. 

Consider an undirected graph $G = (V, E)$, where $V$ is the set of nodes and $E$ is the set of edges. Each node $v_i$ in the graph is associated with a feature vector $\mathbf{x}_i$. The goal of node embedding is to learn a mapping function $f$ that assigns a low-dimensional vector $\mathbf{h}_i$ to each node $v_i$ in the graph:
\begin{equation}
    f : V \rightarrow \mathbb{R}^d, \quad \text{where } \mathbf{h}_i = f(v_i)\label{eq::graph_embed}
\end{equation}
The objective is to minimize a loss function that measures the dissimilarity between the original graph structure and the embedded vectors. One common objective is to preserve the local neighborhood structure, where $\mathcal{N}(v_i)$ represents the neighbors of node $v_i$:
\begin{equation}
\mathcal{L} = \sum_{v_i \in V} \sum_{v_j \in \mathcal{N}(v_i)} \text{sim}(\mathbf{h}_i, \mathbf{h}_j)\label{eq::graph_embed_loss}
\end{equation}
where $\text{sim}(\mathbf{h}_i, \mathbf{h}_j)$ is a similarity function that measures how well the embedding preserves pairwise relationships between nodes.

DeepWalk~\cite{perozzi2014deepwalk} and Node2Vec~\cite{grover2016node2vec} are two common traditional graph embedding methods. DeepWalk assumes nodes that co-occur frequently in random walks are deemed to be semantically similar and thus embeds them closer together in the vector space. Node2Vec extends DeepWalk by introducing flexibility in the random walk strategy, allowing users to control the balance between exploring the local neighborhood and jumping to distant nodes.
Above traditional graph embedding methods provide interpretable and computationally efficient techniques for deriving meaningful representations from graph-structured data. 
To fully encapsulate higher-order dependencies within complex networks, A few works extend beyond pairwise connections to encompass neighbor nodes that are located several hops away~\cite{battiston2020networks}~(See Fig.~\ref{fig:potential}a). Typical ways include modeling local connection patterns such as motifs~\cite{benson2016higher} or utilizing the formation of hypergraphs~\cite{piaggesi2022effective} where each edge can link multiple nodes.

\subsection{Graph neural network} 

Graph Neural Networks (GNNs) represent a powerful advancement in the field of graph representation learning, offering a paradigm shift from traditional graph embedding methods. 
GNNs extend the concept of graph embedding by introducing neural network architectures specifically designed to capture complex relationships and dependencies within graph-structured data.

In specific, GNNs replace the simple mapping function $f$ in Equation (\ref{eq::graph_embed}) with neural networks, which typically comprise linear transformation layers and non-linear activation functions. 
The GNN architecture allows for the stacking of multiple such layers, enabling the acquisition of higher-order information. 
Each layer in this stack aggregates information from neighboring nodes and refines the representations of the nodes in the graph. 
This iterative process ensures that the model captures intricate and higher-level relationships within the graph structure.

Two typical GNN structures that have gained prominence are Graph Convolutional Networks (GCN)~\cite{kipf2016semi} and Graph Attention Networks (GAT)~\cite{velivckovic2017graph}.
GCN forms a foundational architecture in GNNs, in which each node aggregates information from its immediate neighbors, with layer-wise propagation refining node representations. 
Moreover, GAT introduced attention mechanisms to enhance the learning process, which allows nodes to assign different attention weights to different neighbors, enabling the model to focus on the most relevant information during aggregation. 
In contrast to many traditional graph embedding methods, Graph Neural Networks (GNNs) provide a notable advantage through end-to-end training with supervised labels, facilitating the creation of node representations optimized for specific downstream tasks.  
The exceptional capabilities of Graph Neural Networks (GNNs) in capturing intricate patterns that extend beyond immediate neighbors position them as effective tools for tackling the challenge of modeling high-order topological structures within complex networks~(See Fig.~\ref{fig:potential}b). 
Moreover, GNN models seamlessly integrate explicit high-order connections, such as motifs and hypergraphs~\cite{feng2019hypergraph,yadati2019hypergcn}.

\subsection{Dynamic graph learning}

\subsubsection{Dynamic graph structure learning}
The dynamic graph is usually defined as $\mathcal{G}={\mathcal{G}^1, \mathcal{G}^2, \cdots, \mathcal{G}^t}$, where $\mathcal{G}^t = (V^t, E^t)$ of each time step. In this field, a widely recognized problem is to learn the structure of the dynamic graph, \textit{i.e.,} predict further interactions given graph snapshots. Specifically, given two nodes $u$ and $v$ in the time step $t+1$, we intend to learn their embeddings $\bar{h}_u$ and $\bar{h}_v$ based on the temporal features as well as historical snapshots $\mathcal{G}^{1:t}$. Their interaction can be subsequently predicted by a downstream classifier, \textit{i.e.,} $\hat{y}_{uv}^{t+1}=\text{clf}(\bar{h}_u, \bar{h}_v)$, where the ground truth $\hat{y}_{uv}^{t+1}=1$ denotes there is an interaction between $u$ and $v$ at time step $t+1$, otherwise $\hat{y}_{uv}^{t+1}=0$. Real-world dynamic graphs typically exhibit complex spatial-temporal dependencies, preventing direct applications of standard models for static graphs. 
Moreover, many dynamic graphs involve irregularly sampled snapshots, which makes it difficult to capture temporal correlations between consecutive graph snapshots. The widely-used model architecture is the combination of recurrent neural network (RNN) and self-attention mechanism (SAM), while a recent work~\cite{DBLP:conf/iclr/CongZKYWZTM23} finds that a simple architecture based on multi-layer perceptrons and neighbor mean-pooling performs competitively.

\subsubsection{Graph dynamics learning}
Distinct from graph structure learning, graph dynamics learning centers on analyzing and predicting the temporal changes and evolving patterns of node states. 
Consider a dynamic graph $\mathcal{G}^t=(V^t, E^t)$ with $N$ components, $V^t=\{v_1^t, v_2^t, \cdots, v_N^t\}$ denotes node state observations at time step $t$ and $E^t=\{(v_i, v_j)$ denotes their temporal interactions. Given node states and their interactions of $T$ time steps, we aim to predict their states of future $k$ time steps. 
In real-world systems, such as social interactions and pandemic outbreaks, the dynamic nature is not only exhibited in the evolution of node states but also manifests as the changes in graph structure. Towards this end, Huang et al.~\cite{huang2021coupled} utilize coupled Graph-ODE to model the continuous co-evolution of nodes and edges in the latent space, respectively, considering their mutual influence. 

To sum up, data-driven techniques of dynamic graph learning open a new avenue to understanding the complex coevolving dynamics of real-world systems, allowing us to model unknown dynamics accurately from observations without prior knowledge of governing dynamic equations~(See Fig.~\ref{fig:potential}c).

\subsection{Deep generative model} 
Network generation, or graph generation, has profound implications across various fields, revolutionizing our comprehension and interaction with complex systems. It is invaluable in fields ranging from social network analysis to biological systems modeling, offering deep insights into the intricate patterns and dynamics governing these networks.
For example, in social media analysis, generative models can elucidate the dynamics of social interactions and information flow. In drug discovery, they facilitate the identification of new chemical and molecular structures.

A graph can be defined as $G(\mathcal{V},\mathcal{E},X,E)$, where $\mathcal{V}$ is the set of $N$ nodes, $\mathcal{E}\subseteq\mathcal{V}\times\mathcal{V}$ is the set of edges, along with optional node attribute $X$ and edge attribute $E$. $X\in\mathbf{R}^{N\times D}$ is the node attribute matrix with dimension $D$, and $E\in\mathbf{R}^{N\times N \times F}$ is the edge attribute matrix with dimension $F$. The goal of graph generation is to learn the underlying distribution of the observed graphs $\mathcal{G}=\{G_1,G_2,...,G_N\}$ for the purpose of generating new graphs.

Traditional generative models, such as stochastic block models~\cite{holland1983stochastic} and small-world models~\cite{watts1998collective}, have been widely used to generate graphs. However, traditional generative methods are limited in their ability to generalize to real-world networks with complex characteristics and dependencies because they can only model some pre-defined statistical rules of graphs. 
In contrast, deep generative models excel in handling complex network data by leveraging deep learning architectures to capture better nonlinear relationships and intricate patterns from data within networks. By utilizing the distribution modeling capabilities of deep generative models, the expansion and evolution of networks can be effectively captured. This approach bypasses the need for complex rule-based modeling processes in traditional models and provides a detailed representation of both the structural complexity and the diverse information distribution within networks~(See Fig.~\ref{fig:potential}d). 
VAE-based models~\cite{kipf2016variational} focus on encoding graph messages into a latent space, after which the decoder generates new graphs from samples in the latent space. 
GAN-based models~\cite{bojchevski2018netgan} employ a dual-network architecture, one to directly generate graphs and the other to discriminate between realistic and synthetic graphs. 
Autoregressive models~\cite{you2018graphrnn} factorize the graph generation in a sequential way, which determines the next emergence step of nodes and edges upon the previous steps.
However, the lack of explicit co-evolution modeling in VAEs and GANs and the high computational cost of sequential generation only produce networks that are suboptimal in realism while lacking the interpretability of the generation process. 

Recently, diffusion or score-based graph generative models~\cite{niu2020permutation,vignac2022digress} show more scalability by considering the evolution process in a more implicit way, reconstructing the complex dependencies of nodes and edges during the evolution of a complex network structure. 
In conclusion, the development of graph generation approaches marks an important direction toward addressing the co-evolution issue in complex networks, i.e., the dynamic interplay between nodes and their connections over time.

\subsection{Deep reinforcement learning} 
Reinforcement learning~(RL) stands out as a specialized AI paradigm adept at resolving intricate challenges posed by high-dimensional Bellman equation systems. 
These systems effectively describe a range of decision-making and optimization tasks within complex networks.
RL typically operates within the framework of Markov Decision Processes (MDPs), represented as $\langle \mathcal{S},\mathcal{A},P,R, \gamma\rangle$. Here, $s_t\in\mathcal{S}$ denotes the state, $a_t\in\mathcal{A}$ represents the action, $P:\mathcal{S}\times\mathcal{A}\mapsto\mathcal{S}$ characterizes the state transition probability, and $r_t=R:\mathcal{S}\times\mathcal{A}\mapsto\mathbb{R}$ encapsulates the one-step reward.
The goal of RL is to maximize the long-term return, which is calculated in a cumulative discounted manner given the one-step reward as  
$R=\sum_{t=0}^T\gamma^{t}r_t$,
where $\gamma$ is the discounted factor.
In general, an agent interacts with an environment over a series of discrete time steps, making decisions (actions) based on the observed state to maximize cumulative rewards. 
Central to RL are two key concepts: the policy and the value function. 
The policy defines the strategy or behavior of the agent, specifying the probabilities of selecting different actions in a given state. 
On the other hand, the value function evaluates the desirability of states or state-action pairs, quantifying the expected cumulative reward that an agent can attain from a given starting state. 
These components are pivotal in guiding the agent's decision-making process, allowing it to learn optimal strategies for navigating complex environments. 

Deep Reinforcement Learning (DRL) is an extension of traditional RL that leverages the power of deep neural networks to approximate complex policy and value functions~(See Fig.~\ref{fig:potential}e). 
The deep neural networks enable the agent to learn hierarchical and abstract representations of the environment, facilitating more sophisticated decision-making in complex scenarios.
Value-based and actor-critic are two fundamental approaches in DRL, each offering unique strategies for solving complex decision-making problems. 
In the value-based approach, algorithms estimate the value function, quantifying the expected cumulative reward associated with being in a particular state or taking a specific action, as the Q-function approximation in DQN~\cite{mnih2013playing}.
On the other hand, the actor-critic approach combines aspects of both policy-based and value-based methods. 
The actor, represented by a policy network, decides which actions to take in a given state, while the critic, usually a value network, evaluates the desirability of those actions. 
Proximal Policy Optimization (PPO)~\cite{schulman2017proximal} is a typical example of an actor-critic algorithm that introduces a clipped surrogate objective, preventing large policy updates that could lead to instability.

\subsection{Physics-informed machine learning} 
Physics-informed machine learning (PIML) integrates physical laws into machine learning models, enhancing their ability to make predictions consistent with known physical principles. 
In complex network scenarios, this approach becomes particularly crucial, given the intrinsic characteristics and challenges associated with network data.
On the one hand, networks represented as geometric graphs often exhibit symmetries of translation and rotation. For instance, in molecular graphs, each atom might be represented by both scalar attributes (like charge) and geometric attributes (like position), and outputs of models should remain consistent under rotational transformations when applied in predicting molecular dynamics. on the other hand, the dynamic evolution of graphs sometimes follows the guiding principles of physical constraints. Incorporating geometric symmetries and physical rules into graph learning helps improve accuracy and generalization, preventing overfitting and allowing for training with imperfect data by adhering to expected physical behaviors.

To embed machine learning models with physics information, a method for integrating biases should be properly devised. One important approach adopted in PIML models is to involve inductive biases in the architecture~\cite{karniadakis2021physics}.
In general graph deep learning, these biases often manifest as permutation invariance and geometric equivariance, crucial for accurately processing graph-structured data. 
For example, the Equivariant Graph Neural Network (EGNN)\cite{satorras2021n} explicitly models the equivariance of graph data under Euclidean transformations.
Moreover, learning biases, introduced through specific loss functions, are significant in guiding models toward solutions that obey physical principles. This approach allows for a flexible yet controlled physics integration into the learning process. 
Zheng et al.\cite{zheng2023towards} demonstrated using energy function-based loss methods to guide pre-training, effectively addressing the challenge of limited training data while ensuring a physics-informed learning process.
By embedding physical principles directly into the learning architecture and loss functions, these models achieve a unique balance between data-driven insights and fundamental physical understanding. 
Similarly, Chen et al.\cite{chen2024social} also leveraged social physics knowledge to better simulate human crowd movements across a wide range of realistic environments.

\section{Research Problems and Methodology} \label{sec:pam}
\begin{figure}[t]
    \centering
    \hspace{-2em}
    \includegraphics[width=1.05\linewidth]{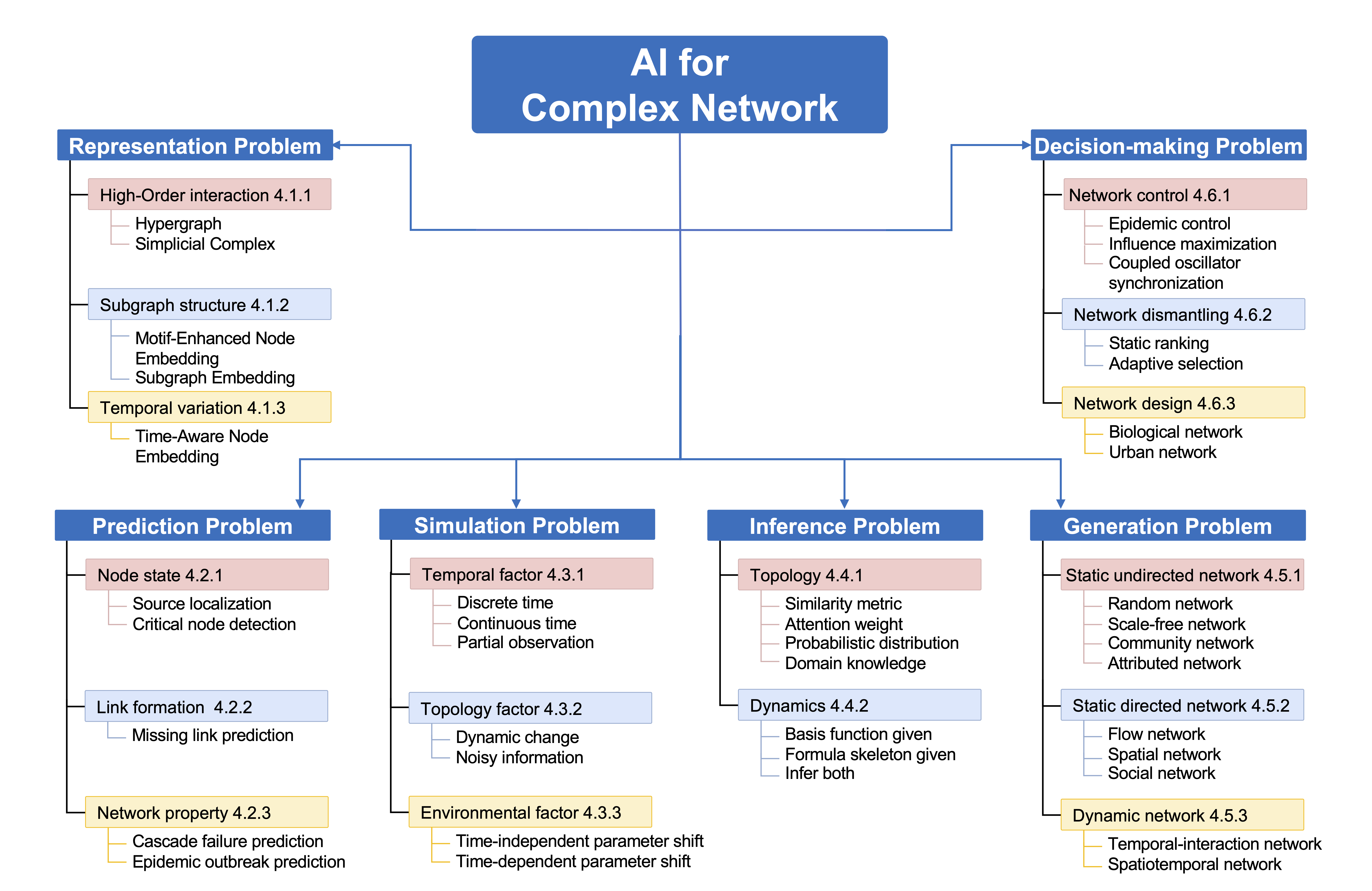}
    \vspace{-2em}
    \caption{Taxonomy of research problems in AI for complex network.}
    \vspace{-1em}
    \label{fig:problem}
\end{figure}

In this section, we summarize key research problems of AI for complex networks into six categories, covering the representation, prediction, simulation, inference, generation and control~(or decision-making) of complex networks. The detailed taxonomy is illustrated in Fig.~\ref{fig:problem}. Meanwhile, we also discuss AI models designed for different problems and make a comprehensive comparison in Table~\ref{tab:method}.

\begin{table}[t]
\small
\setlength\tabcolsep{1pt}
\renewcommand{\arraystretch}{1.05}
\caption{Details of AI models designed for different problems in complex networks.}
\vspace{-1em}
\begin{tabular}{cccc}
\toprule
\textbf{Problem} & \textbf{Category} & \textbf{Methodology} & \textbf{Model}\\
\toprule
\multirow{7}{*}{Representation} & \multirow{3}{*}{\makecell[c]{High-Order\\ interaction}} & Hyperedge embedding & Hyper-gram~\cite{huang2019hyper}, SIMPLEX2VEC~\cite{piaggesi2022effective} \\
& & Hypergraph NN & HGNN~\cite{feng2019hypergraph}, HNN~\cite{srinivasan2021learning} \\
& & Hodge Laplacian + GNN & Hodge1Lap~\cite{zhou2023facilitating} \\ \cline{2-4} 
& \multirow{3}{*}{\makecell[c]{Subgraph\\ structure}} & Motif matrix decomposition & HONE~\cite{rossi2020structural} \\
& & Motif matrix + GNN & MTSN~\cite{liu2021motif} \\ 
& & GNN on Motif-Motif Network & MHM-GNN~\cite{cotta2020unsupervised}, HM-GNN~\cite{yu2022molecular} \\ \cline{2-4} 
& \makecell{Temporal \\ variation} & Temporal Shift GNN & \makecell{MTSN~\cite{liu2021motif}, TREND~\cite{wen2022trend}, \\ SpikeNet~\cite{li2023scaling}} \\\midrule

\multirow{5}{*}{Prediction} & \multirow{2}{*}{\makecell[c]{Node\\ state}} & {GNN} & \makecell{ILGR~\cite{munikoti2022scalable}, HONNMA~\cite{zhao2023novel}, \\ Deepinf~\cite{qiu2018deepinf}, GCNSI~\cite{dong2019multiple}, IVGD~\cite{wang2022invertible}} \\
& & {Deep generative model} & DDMSL~\cite{yan2023diffusion}, SL-Diff~\cite{huang2023two} \\\cline{2-4} 
& \multirow{2}{*}{\makecell[c]{Link\\ formation}} & {Graph embedding} & {GraphSAGE~\cite{hamilton2017inductive}, NetMF~\cite{qiu2018network,qiu2019netsmf}} \\
& & {GNN on local subgraph} & {WLNM~\cite{zhang2017Weisfeiler}, SEAL~\cite{zhang2018link},  NCN~\cite{wang2023neural}} \\\cline{2-4} 
& Network property & GNN + classifier          & \makecell{CF-GCN~\cite{ahmad2021predicting}, LRP-GCN~\cite{liu2021searching}, \\ MAML-MPNN~\cite{panagopoulos2021transfer}, CausalGNN~\cite{wang2022causalgnn}} \\\midrule

\multirow{6}{*}{Simulation} & \multirow{3}{*}{\makecell[c]{Temporal\\ factor}}  & GNN + RNN & G-RNN~\cite{murphy2021deep} \\
& & Graph ODE & G-ODE~\cite{zang2020neural}, GP-ODE~\cite{yildiz2022learning}, HOPE~\cite{luo2023hope} \\
& & Transformer & AgentFormer~\cite{yuan2021agentformer} \\\cline{2-4} 
& \multirow{2}{*}{\makecell[c]{Topology\\ factor}} & Graph ODE & CG-ODE~\cite{huang2021coupled} \\
& & Spatio-temporal GAT & DIDA~\cite{zhang2022dynamic} \\\cline{2-4} 
& Environ. factor & Disentangled Graph ODE & GG-ODE~\cite{huang2023generalizing}, CARE~\cite{luo2023care} \\\midrule

\multirow{5}{*}{Inference} &\multirow{3}{*}{Topology} & Metric Learning + GNN & {IDGL~\cite{chen2020iterative}, dDGM-RG~\cite{de2022latent}, Grale~\cite{halcrow2020grale}} \\
& & Attention-mechanism + GNN & RAIN~\cite{ha2022learning}, IMMA~\cite{sun2022interaction} \\\
& & Graph generator + GNN & Slaps~\cite{fatemi2021slaps}, dDGM~\cite{kazi2022differentiable}, NRI~\cite{kipf2018neural} \\\cline{2-4} 
& \multirow{2}{*}{Dynamics} & SINDy & G-SINDy~\cite{gao2022autonomous} \\
& & Symbolic regression + GNN & SymDL~\cite{cranmer2020discovering}, GraphSR~\cite{shi2022learning} \\\midrule

\multirow{7}{*}{Generation} & \multirow{3}{*}{\makecell[c]{Static \\ undirected \\ network}} & Autoregressive  & GRAN~\cite{liao2019efficient}, GraphRNN~\cite{you2018graphrnn} \\ 
& & GAN             & SPECTRE~\cite{martinkus2022spectre} \\ 
& & Diffusion model       & DiGress~\cite{vignac2022digress}, EDGE~\cite{DBLP:conf/icml/ChenHH023} \\ \cline{2-4} 
& \multirow{2}{*}{\makecell[c]{Static directed \\ network}}   & Edge rewiring                 & EdgeRewire~\cite{schweimer2022generating} \\ 
& & GAN             & FlowGEN~\cite{kocayusufoglu2022flowgen} \\ \cline{2-4} 
& \multirow{2}{*}{\makecell[c]{Dynamic \\ network}}           & GAN & TagGen~\cite{zhou2020data} \\ 
& & VAE            & SGD-VAE~\cite{guo2021deep}, STGD~\cite{du2022disentangled} \\ \midrule

\multirow{7}{*}{\makecell[c]{Decsion-making \\ Control}} & \multirow{3}{*}{\makecell[c]{Network \\ control}}   & Constrained minimization & DCCN~\cite{baggio2021data} \\
& & GNN + RL & \makecell{RL4IM~\cite{chen2021contingency}, RLGN~\cite{meirom2021controlling}, DeepIM~\cite{ling2023deep}, \\  MoMRL~\cite{wan2021multi},RL4Aging\cite{sun2020optimal}}\\
& & Neural ODE & NODEC~\cite{bottcher2022ai,asikis2022neural} \\\cline{2-4} 
& \multirow{2}{*}{\makecell[c]{Network \\ dismantling}}  & GNN & GDM~\cite{grassia2021machine}, NIRM~\cite{zhang2022dismantling} \\
& & GNN + RL & FINDER~\cite{fan2020finding} \\\cline{2-4} 
& \multirow{2}{*}{\makecell[c]{Network \\ design}}  & One-shot & {STransformer~\cite{ingraham2019generative}, ProteinSGM~\cite{lee2023score}}\\
& & Step-by-step & DIRAC~\cite{fan2023searching}, DRL4Urban~\cite{zheng2023spatial,zheng2023road} \\\bottomrule

\end{tabular}%
\label{tab:method}
\vspace{-1em}
\end{table}

\subsection{Representation problem}

Representation learning in complex networks aims to transform the unstructured networked data into a structured low-dimensional vector space, so that similar or neighboring entities (e.g., nodes, motifs, edges, and even networks themselves) are close to each other, while dissimilar entities are far apart. Representation learning in complex networks is a fundamental task in the field of complex networks, which can be applied to many downstream tasks, such as node classification, link prediction, and network visualization. In this section, we will focus on the unique challenges of representation learning in complex networks compared to ordinary graphs, including high-order interaction, subgraph structure, and temporal variation, as well as the role of AI methodology in addressing these challenges. The related methods are listed in Table~\ref{tab:method}.

\subsubsection{High-order interaction}
High-order interactions are widely present in complex networks, such as multi-protein complexes in protein-protein interaction networks, communities in social networks, and multi-author co-authored papers in citation networks. 
Researchers often give up using the standard graph that can only represent pairwise interactions between nodes but use some more complex representation methods, such as \emph{hypergraphs} or \emph{simplicial complexes}. Due to space limits, we only discuss hypergraphs here and leave simplicial complexes in Appendix~\ref{app:m1}. 

A hypergraph is an extension of the standard graph, which provides the most general and unconstrained description of high-order interactions\cite{battiston2020networks}. Formally, a hypergraph includes a set of nodes $\mathcal{V}$ and a set of hyperedges $\mathcal{H}$, where each hyperedge $h \in \mathcal{H}$ is a subset of nodes $h \subseteq \mathcal{V}$ and represents a high-order interaction among the nodes in $h$. 
Inspired by Word2Vec\cite{mikolov2013distributed} and DeepWalk\cite{perozzi2014deepwalk}, Huang et al.\cite{huang2019hyper} defined a concept called hyper-path and designed hyper-path-based random walks that can learn representation of hypergraph while preserving its structural information.
As a comparison, Feng et al.\cite{feng2019hypergraph} introduced GNN to learn the representation of hypergraphs. Specifically, they modeled the hypergraph by an incidence matrix 
$\mathbf{M} \in \mathbb{R}^{|\mathcal{V}|\times|\mathcal{H}|}$, where $M_{ij} = \mathbb{I}(v_i \in h_j), \forall v_i \in \mathcal{V}, h_j \in \mathcal{H}$, 
and designed a hyperedge convolution layer that performs node-hyperedge-node message passing to refine the node features on the hypergraph.
However, this method cannot capture the implicit higher-order interactions between the nodes/hyperedges, so Srinivasan et al.\cite{srinivasan2021learning} used a message-passing framework that explicitly receives messages from not only the hyperedges it belongs to but also neighboring vertices in the two-hop neighborhood.

\subsubsection{Subgraph structure}
\label{sec:4.1.2}
In real-world complex networks, some patterns of subgraph structures are significantly more frequent than those in randomized networks, such as triangles, 4-vertex rings, and 4-vertex fully connected subgraphs\cite{milo2002network}. These subgraph structures are called motifs. Motif is another way to describe high-order interactions in complex networks. Unlike hypergraphs or simplicial complexes, which relax the condition that an edge should link exactly two nodes, motifs describe high-order interactions within the standard pairwise graph, and thus offer distinct advantages in scenarios where directly modeling high-order interactions is challenging or where simplicity in graph model is preferred.
Some work leverages motifs to capture higher-order structural dependencies and connectivity patterns in complex networks and thus help to learn representations for nodes considering the subgraph structural similarity. Others proposed learning the representation of motifs themselves for some specific purpose.

For the works that learn motif-enhanced node embedding,
Rossi et al.\cite{rossi2020structural} proposed to learn node representation with motif matrix indicating co-occurrence of each node pair within a certain type of motif.
Liu et al.\cite{liu2021motif} used motif matrix as well, but in contrast, they designed a new message-passing matrix in GCN based on the motif matrixes to learn the node representation that preserves the local high-order structures.
For the works that learn embedding for motifs,
Cotta et al.\cite{cotta2020unsupervised} proposed a GNN-based method to learn representations for motifs to get joint representations of nodes.
On the other hand, Yu et al.\cite{yu2022molecular} focused on learning representations for chemical molecules while still adopting a GNN-based method to learn representations for motifs. Please refer to Appendix~\ref{app:m1} for more methodology details.

\subsubsection{Temporal variation}
Complex networks that consist of the temporal activity of nodes and edges, such as population evolution of different species in ecosystems and dynamic neural activity in brain networks, are called dynamic networks or temporal networks. In these networks, representation learning must consider the temporal dynamics of nodes and edges to capture the dynamic evolution of the network.
Therefore, researchers usually focus on incorporating the temporal dynamics of nodes and edges in different time snapshots into the model to get time-aware node representation. Please refer to Appendix~\ref{app:m1} for more methodology details.

\subsubsection{Future direction}
AI approaches exhibit significant advantages over traditional methods in learning representation for complex networks. A promising direction is to design \textbf{Dynamics-aware representation for high-order network}. High-order interactions and dynamics are two common properties that are widely observed in real-world complex networks.
However, we found that there is little work that considers the dynamic and high-order interactions of complex networks at the same time, which can be well handled by neural networks with well-designed structures.

\subsection{Prediction problem}

Prediction is another fundamental task in complex networks due to its practical value in various applications. According to the entities to be predicted, the prediction task can be divided into three categories: 1) node state prediction, such as source localization and critical node detection, 2) link formation prediction, such as missing link prediction, and 3) network property prediction, such as cascade failure prediction and epidemic outbreak prediction.

\subsubsection{Node state prediction}
\emph{Source Localization} tasks aim to retrieve the initial source node in the case of knowing the network topology and the future network cascade pattern, which has a wide range of applications, such as rumor detection, malware tracing, and disease propagation source localization. Formally, considering a network with $N$ nodes, a source node vector $x \in \{0, 1\}^N$ represents whether a node is infected at the initial timestep. Infected nodes at time $t-1$ can infect some of their neighboring nodes in the next time step $t$ and thus cause a new infected vector $y_t \in \{0, 1\}^N$. The source localization task aims to retrieve the source node vector $x$ from the diffusion vector $y_T$ and network topology $A$.
Instead of localizing sources directly~\cite{dong2019multiple}, most works proposed to address it as an inverse problem of graph diffusion~\cite{wang2022invertible,yan2023diffusion,huang2023two}. 
For example, Wang et al.\cite{wang2022invertible} introduced a graph diffusion model that, after training, can be inversed with two fixed point iterations to get a source node estimation from the diffusion vector.
Yan et al.\cite{yan2023diffusion} also considered the inverse problem of graph diffusion and adopted a generative diffusion model to solve it.
\emph{critical node detection} aims to find the key nodes that determine the system output more critically than others, with a wide range of applications such as maximizing influence in social networks, controlling epidemic spreading, and protecting species in ecosystems. Please refer to Appendix~\ref{app:m2} for more methodology details.

\subsubsection{Link formation prediction}
The major task in link formation prediction is \emph{missing link prediction}, which aims to predict the unknown link in the network under the known partial network topology, which has a wide range of applications in the fields of protein network prediction, drug-drug interaction prediction, etc.
Apart from these graph embedding-based methods~\cite{hamilton2017inductive,qiu2018network,qiu2019netsmf}, other works predict whether two nodes in a network are likely to have a link by considering the local subgraph of their $h$-hop neighbors. For example,
to predict the link between nodes $(u, v)$, Zhang et al.\cite{zhang2017Weisfeiler} proposed to extract the subgraph induced by their $h$-hop neighbors and then apply an MLP to the adjacency matrix of this subgraph to predict the likelihood of the link between them.
Zhang and Chen\cite{zhang2018link} replaced the MLP with GNNs to empower it to handle local subgraphs with different sizes and learn from node features as well as an adjacency matrix.
Wang et al.\cite{wang2023neural} proposed a Neo-GNN method inspired by the classical Common Neighbour~(CN) method. In contrast to CN, Neo-GNN considers multi-hop common neighbors and introduces a learnable function to improve performance. Please refer to Appendix~\ref{app:m2} for more methodology details.

\subsubsection{Network property prediction}
\emph{Cascade failure prediction} aims to detect the early onset of a cascading failure, which means a quick succession of multiple component failures that take down the system. This phenomenon is widely observed and studied in many real-world complex networks, especially the power system.
For example, Ahmad et al.\cite{ahmad2021predicting} introduced a spatio-temporal GCN to predict the potential cascading failure in power systems. 
Liu et al.\cite{liu2021searching} also focused on the cascade failure in power systems while they aimed to identify as many critical cascading failures as possible for a given network. 
\emph{Epidemic outbreak prediction} is another important task in complex network prediction. It aims to predict the potential outbreak of an epidemic in a networked system, which has a wide range of applications in the fields of disease control, public health, etc.
Panagopoulos et al.\cite{panagopoulos2021transfer} proposed a GNN-based model to predict the outbreak of COVID-19 in different countries. They also leveraged Model-Agnostic Meta-Learning (MAML) to make their model generalizable among different countries.
Wang et al.\cite{wang2022causalgnn} focused on a similar problem and adopted a GNN-based model as well. However, they further considered the underlying causal mechanisms to solve the challenge that epidemic data is sparse and noisy and a deep learning model thus tends to overfit. Please refer to Appendix~\ref{app:m2} for more methodology details.

\subsubsection{Future directions}

AI approaches exhibit significant advantages over traditional methods in predicting node states, link formation, and network properties for complex networks. Promising directions include:
\begin{itemize}
    \item \textbf{High-order structural information in source localization}. We noticed that the graph diffusion typically involves local subgraph structures.
    Therefore, it is reasonable to consider the role of motifs in enhancing source identification accuracy. 
    \item \textbf{Casecade failure prediction in more complex networks}. Most existing works focused on the power system and adopted corresponding domain knowledge in their modeling, which makes it difficult to generalize to other types of real-world complex networks, such as the Internet, transportation networks, and ecological networks. 
    \item \textbf{Prediction of macroscopic properties related to network dynamics}. The prediction of macroscopic properties related to network dynamics, such as robustness and resilience~\cite{artime2024robustness}, are important for understanding the mechanisms of complex networks. 
    Although some classical methods for predicting these properties exist~\cite{ross2021universal}, they often suffer from high computational complexity and thus cannot deal with large-scale complex networks. 
\end{itemize}

\subsection{Simulation problem}

 The accurate simulation of real-world network dynamics is the critical step in understanding such complex systems and predicting their future state. 
 Given the graph $\mathcal{G}=\{\mathcal{V},A\}$, where $\mathcal{V}=\{v_i\}^N_{i=1}$ is the vertex set and $A=(a_{ij})$ is the adjacency matrix, we denote node feature matrix as $X=\{x_i\}^N_{i=1} \in \mathbb{R}^{N \times d}$ and the system dynamics $\mathcal{M}$ as
$\{X_{t+\Delta t},\mathcal{G}_{t+\Delta t}\}=\mathcal{M}(X_t,\mathcal{G}_t)$,
where $(\cdot)_t$ means the value at time $t$. The objective of the simulation is to build a model $\hat {\mathcal{M}}$ parametrized by a set of parameters $\theta$ from the observed data, which satisfies
$\hat {\mathcal{M}}(X_t,\mathcal{G}_t, \theta) \approx \mathcal{M}(X_t,\mathcal{G}_t)$.
When the graph topology is fixed, the problem can be simplified as
$X_{t+\Delta t}=\hat {\mathcal{M}}(X_t, \mathcal{G}, \theta)$.
The simulation of complex networks confronts multiple challenges as data availability deteriorates, incorrect topologies, and environmental variations.

\subsubsection{Temporal factor.} 

The temporal factor involves the simulation in both discrete and continuous time, taking into account the constraints of partial observation. In the context of discrete time, Murphy et al.~\cite{murphy2021deep} designed a GNN-based deep learning model to successfully model the dynamics of contagion networks.  Yuan et al.~\cite{yuan2021agentformer} propose a new Transformer, AgentFormer, that simultaneously models the time and social dimensions to predicte accurate future trajectories of multiple agents. 
Continuous-time dynamics on complex networks are mostly modeled by the Graph ODE. For example, Zang et al.~\cite{zang2020neural} combine Ordinary Differential Equation Systems~(ODEs) and GNNs to learn continuous-time dynamics on complex networks in a data-driven manner. By employing latent Gaussian process ordinary differential equations, Yıldızet al.~\cite{yildiz2022learning} infer both independent dynamics and their interactions with reliable uncertainty estimates. Luo et al.~\cite{luo2023hope} utilize a second-order graph ODE function that models the dynamics for both nodes and edges in the latent space respectively.
In addition, partial observation frequently occurs in real-world systems, where observations may not be temporally aligned for different nodes~\cite{huang2020learning}.  Please refer to Appendix~\ref{app:m3} for more methodology details.

\subsubsection{Topology factor.}
Dynamic topology is also common in real-world networks, and the simulation task degenerates into: Given an initial graph $\mathcal{G*}=\{\mathcal{V}, A*\}$, how to build a model $\hat {\mathcal{M}}$ which can model the evolution of topology.
Coupled Graph ODE~\cite{huang2021coupled} models the continuous evolution of nodes and edges in the latent space by two coupled ODE functions, respectively. DIDA~\cite{zhang2022dynamic} captures the variant and invariant patterns by a disentangled spatio-temporal attention network, which helps to handle spatio-temporal distribution shifts in dynamic graphs. As for noisy information that exists in observed topology~\cite{yang2019topology,peng2022reverse}, Please refer to Appendix~\ref{app:m3} for more methodology details.

\subsubsection{Environmental factor.}
The environmental variations also affect the network dynamics, indicated as time-independent parameter shift~\cite{huang2023generalizing} or time-independent parameter shift~\cite{luo2023care}. The dynamics of such complex networks can be summarised as 
$\{X_{t+\Delta t}, \mathcal{G}_{t+\Delta t}\}=\mathcal{M}(X_t,\mathcal{G}_t,E)$,
where $E$ is either a finite discrete value for the switch of environments or a time-varying variable for the continuous variations in the same environment. The exogenous factors $E$ can span over a wide range of settings such as pressure, gravity, and temperature, which influence the system dynamics, making the simulation more challenging. 

Considering the modeling of dynamics governed by common physical laws in different environments, namely time-independent parameter shift, GG-ODE~\cite{huang2023generalizing} trains a shared Graph ODE guided by the thought of minimizing mutual information to design an environmental encoder that decouples environmental features and general dynamics from the data. 
For time-dependent parameter shift, CARE~\cite{luo2023care} separates contextual variables of environmental fluctuations and pure node features from historical trajectories, which are then used together in Graph ODE to model network dynamics.

\subsubsection{Future Directions.} The current AI methods are able to accelerate the simulation of complex networks without assumptions on dynamical equations. The future directions are:
\begin{itemize}
    \item \textbf{Incomplete topology.} data-driven simulation approaches are less robust under incorrect or incomplete topology that is common in practice~\cite{jiang2020true, yang2019topology}. Extracting accurate topology through end-to-end optimization of downstream tasks is a potential direction.~\cite{chen2020iterative, peng2022reverse, huo2023t2}.
    \item \textbf{OOD generalization.} Real-world applications call for the need to design more versatile models for simulating complex networks under a time-shift environment.
    \item \textbf{Multi-scale simulation.} Complex networks often exhibit different behavioral and structural characteristics at different scales~\cite{vlachas2022multiscale}. 
    Therefore, it is necessary to separate different scales and model them independently~\cite{li2023learning2}. 
\end{itemize}

\subsection{Inference problem}

Complex networks in the real world span biological systems and social organizations, and the dynamics of such real networks are often unclear. Given the partial knowledge of the network mechanisms, data-driven AI technologies have the potential to infer complex network topology and dynamics from data automatically.

\subsubsection{Topology Inference} Since the topology of a real-world complex network is often unknown or partially known, inferring possible connectivity from the historically observed states of nodes is essential for understanding the mechanisms of complex networks. 
We denote the node state matrix as $X=\{x_i\}^N_{i=1} \in \mathbb{R}^{N \times F}$, where $x_i \in \mathbb{R}^{1 \times d}$ is the $d$-dimensional state vector of node $v_i$. Given the historical node states $\{X_1, X_2, ... , X_n\}$ or labels $\mathcal{C}=\{c_i\}^N_{i=1}$, the goal of topology inference is to infer the surrogate adjacency matrix $\overline A$ from the dynamics of the network, which enables optimal prediction of future states or classification of nodes. Existing studies can be categorized into the following four types of approaches.

\textbf{Similarity metric} based approaches generate edge weights according to similarity between pairwise node representations, such as the Gaussian kernel~\cite{wu2018quest} and cosine similarity~\cite{chen2020iterative}. Please refer to  Appendix~\ref{app:m4} for more methodology details on computationally efficient metrics~\cite{cosmo2020latent,de2022latent}, fused similarity metrics~\cite{halcrow2020grale} and sparsity-guaranteed metrics~\cite{wu2018quest, halcrow2020grale}.
\textbf{Attention weight} based approaches adopt graph attention mechanism~\cite{velivckovic2017graph} to provide the interaction strength of each pair of nodes as the edge weights. RAIN~\cite{ha2022learning} used LSTM to encode each agent's trajectory and aggregated the hidden states from other agents according to the interaction strengths identified by the graph attention module, which is used to predict future trajectories. To ensure interpretability, multiplex attention in IMMA~\cite{sun2022interaction} learns one interaction type at a time, inferring the type of connection between agents incrementally.
\textbf{Probabilistic distribution}  based approaches used learnable parameters~\cite{franceschi2019learning, fatemi2021slaps} or generative models~\cite{kazi2022differentiable, kipf2018neural} to sample the topology. dDGM~\cite{kazi2022differentiable} used the Gumbel softmax trick to sample a sparse k-degree graph.
Please refer to  Appendix~\ref{app:m4} for more methodology details on \textbf{domain knowledge} based approaches.

\subsubsection{Dynamics Inference} The discovery of governing equations from observed data can help us understand and predict the behavior of complex networks. 
Gao et al.~\cite{gao2022autonomous} designed a two-phase approach based on SINDy to infer the equations of the self-dynamics and the interaction dynamics under the condition where the topology $A$ is known. 
Cranmer et al.~\cite{cranmer2020discovering} relax the assumption of topology and train GNNs to model complex network dynamics in a supervised manner, followed by the symbolic regression~(SR) to discover explicit self-dynamics and interaction dynamics of nodes. This approach both leverages the strong inductive biases of GNNs for complex network data and uncovers the black box of deep learning with the help of SR.
Furthermore, Shi et al.~\cite{shi2022learning} generalize the problem of learning formulas with given skeletons~(inductive bias) from graph data in \cite{cranmer2020discovering} by additionally learning the formula skeleton from data.

\subsubsection{Future direction} 
In terms of complex network inference, future directions include:
\begin{itemize}
    \item \textbf{Inferring topology and dynamics simultaneously.} Existing works on dynamics inference ~\cite{cranmer2020discovering, shi2022learning} are based on the assumption of fully connected graphs, which means that the computational overhead will be unaffordable for a larger network. Therefore, a comprehensive framework that can simultaneously infer sparse adjacency matrices and node dynamics is a potential direction.
    \item \textbf{High-order correlation.}  The inference of high-order correlations in network dynamics still lacks relevant research. Inferring the interactions among multiple nodes beyond pairwise relations is particularly beneficial for understanding the underlying mechanisms of complex network dynamics. 
\end{itemize}

\subsection{Generation problem}\label{subSec: Generation}

Complex network generation involves the development of generative methods to create novel networks that can not only exhibit complex structures and properties but also be similar to those observed in the real world. 
In this section, we categorize common networks into three types: static undirected, static directed, and dynamic networks. 

\subsubsection{Static undirected network.}
The generation of typical static undirected networks, including random networks, scale-free networks, and community networks, is predominantly driven by underlying mechanisms such as random connectivity, preferential connectivity, etc. To capture such mechanisms automatically, researchers have designed effective AI approaches that are capable of modeling the high-dimensional nature of networks and complex, non-local dependencies existing between edges. Please refer to Appendix~\ref{app:m5} for more methodology details on deep generative modeling approaches designed for not only generating pure networks with complex topology but also with attributes corresponding to various real-world entities such as proteins, molecules, transportation systems, and ecological systems.

\subsubsection{Static directed network.}
Directed networks, distinct from their undirected counterparts due to the asymmetry in their connections, play a crucial role in accurately representing phenomena where the direction of interaction is a critical aspect of the network structure and dynamics, such as the real-world scenario of transportation flow networks, spatial networks, social networks, etc.
Kocayusufoglu et al.~\cite{kocayusufoglu2022flowgen} focus on generating realistic directed flow networks to facilitate applications in urban infrastructure design, transportation, and social sciences. The authors propose FlowGEN, a GAN-based framework designed to generate both the directed flow network topology and numerical flows within it. The framework consists of a permutation invariant flow-pooling layer, bi-directional message-passing layers, and an attention-based readout layer, all contributing to learning hidden node representations that capture the complex interplay between network topology and its numerical flows. 
Guo et al.~\cite{guo2021deep} propose to synthesize spatial networks via a spatial-network variational autoencoder (SND-VAE), along with a new spatial-network message passing to discover dependent and independent factors of spatial-level and network-level factors. The authors also incorporate a novel objective for joint spatial-network factor disentanglement, which not only benefits generation performance but also enhances model interpretability.
Another work conducted by Schweimer et al.~\cite{schweimer2022generating} aims to generate directed social networks to be used as surrogates in social science analysis. The authors sample the same number of nodes from the real network and manipulate the density and average clustering coefficient in the largest weakly connected component of the generated networks by edge rewiring to create networks that closely resemble the real ones.

\subsubsection{Dynamic network.}
Many real-world networks are actually dynamic in nature and represented as a series of network snapshots, which capture the evolution and temporal changes of the network structure over time. Dynamic networks usually consist of temporal interaction networks and spatiotemporal networks.
A recent work of Zhou et al.~\cite{zhou2020data} concentrates on preserving both the structural and temporal characteristics of the real data and synthesizing realistic dynamic networks. The proposed TagGen model addresses challenges in dynamic network generation by combining a random walk sampling strategy with a bi-level self-attention mechanism, which efficiently captures structural and temporal context from input graphs. Moreover, TagGen utilizes a network context generation scheme for node and edge addition and deletion, simulating dynamic system evolution, with a discriminator module ensuring the plausibility of these operations.
Du et al.~\cite{du2022disentangled} aim to generate spatiotemporal networks based on a Bayesian model, which factorizes spatiotemporal networks into spatial, temporal, and network factors as well as models factors that explain the interplay among them. 
Temporal networks, 

\subsubsection{Future Directions.}
Several types of special networks that lack generation approaches include: 
\begin{itemize}
    \item \textbf{Dynamic network generation.} There is a need for more powerful models that can handle increasingly complex network structures, particularly in the realm of dynamic networks. While significant progress has been made, the full potential of dynamic network generation remains untapped, with opportunities to better capture the nuances of network evolution and high-order interactions~\cite{do2020structural,longa2024generating}.
    \item \textbf{Weighted network generation.} Weighted networks are common in real-world scenarios such as ecological networks (where edges represent mutualistic or competitive strength) and transportation networks (where edges represent distances or travel times). 
    \item \textbf{Multi-attributed network generation.} The incorporation of both network structure and multiple attributes imparts semantics to nodes and edges, rendering synthetic networks more authentic and versatile. For example, in a transportation network, nodes may depict cities with attributes such as population, while edges might encompass details like transportation mode (e.g., road, rail, air) and capacity. 
\end{itemize}

\subsection{Decision-making and control problem}

Decision-making and control in complex networks aim to intervene in the nodes or edges to influence the behavior of the entire complex network.
Due to the catastrophic dimensions of complex networks, it is extremely challenging to obtain the optimal decision solutions within affordable time.
In this part, we focus on several typical network decision-making tasks for which AI approaches have been employed, including network control, network dismantling, and network design.

\subsubsection{Network control}
The ability to control the network is the ultimate goal in understanding and studying complex networks.
Since complex networks are a coupling of topology and dynamics, network control requires a thorough consideration of both aspects to make the network reach the desired state.
Notably, the controlling agent typically lacks knowledge of the underlying network dynamics, whereas the state trajectories, being more readily accessible, are commonly adopted as input for network control algorithms.

Existing methods can be divided into three main categories.
First, using the trajectories, closed-form solutions can be obtained by solving a constrained minimization problem under linear assumptions, and the optimal control input is derived to steer the system from the initial state to the desired state~\cite{baggio2021data}.
Second, RL approaches were proposed which take the network trajectories as states and learn the mapping from the observed states to the optimal actions.
For example, Meirom \textit{et al}~\cite{meirom2021controlling} transformed the observed network dynamics as a temporal graph, and proposed a GNN model considering both local diffusion and long-range information correlation.
To control the network, $k$ nodes are selected according to the estimated score from the learned node embeddings by GNN.
The authors apply the GNN based RL model to control diffusive processes on complex networks, and experiments on epidemic control showcase the potential of RL and GNN in controlling network dynamics.
Multi-objective model-based RL models~\cite{wan2021multi} were also proposed for infectious disease control to find Pareto-optimal policies that make trade-offs between controlling the infected nodes and reducing the economic costs.
Meanwhile, Sun \textit{et al}~\cite{sun2020optimal} combined optimal control theory and reinforcement learning to determine maintenance protocols for aging networks.
Specifically, repairing schedules are synthesized to promote the longevity of the system at minimal intervention cost.
Third, the neural ODE model is employed to generate control actions, driving the trajectories of the network to the desired states~\cite{asikis2022neural, bottcher2022ai}.
For example, Asikis \textit{et al}~\cite{asikis2022neural} utilized neural ODE to generate control signals with low energy.
Please refer to Appendix~\ref{app:m6} for more methodology details on typical network control scenarios.

\subsubsection{Network dismantling}

Network dismantling involves strategically identifying and removing a minimal set of critical nodes or links to disrupt the overall functionality of a network. 
This task is particularly relevant in various fields such as transportation, communication, and social networks, where understanding the vulnerabilities and weaknesses of a network can have significant implications. 
The goal of network dismantling is to unveil the structural vulnerabilities of a system, providing insights into its robustness and resilience. 
Current AI methods fall into two main categories.
First, graph neural networks, especially graph attention networks~\cite{velivckovic2017graph}, are leveraged to learn representations for nodes and rank them to determine the attack order~\cite{grassia2021machine}.
Supervised learning data are constructed from the original graph and the dismantled graph, and the GNN is trained to classify whether each node belongs to the dismantled graph.
Zhang \textit{et al}~\cite{zhang2022dismantling} further develop both local and global scoring modules and fuse these two aspects to compute the overall contribution of each node.
Second, as an NP-hard problem with an extremely large solution space, RL algorithms demonstrate notable advantages in efficiently exploring potential solutions.
For example, Fan \textit{et al}~\cite{fan2020finding} leveraged Deep Q-Learning~\cite{mnih2015human}, a widely adopted reinforcement learning algorithm, to estimate the value of each node based on the learned node embeddings from GNN and identify critical nodes.
The model shows superior generalization ability, as it was trained in extensive sets of small synthetic networks, and it outperformed existing approaches on large-scale real-world networks.

\subsubsection{Network Design}
Network search and design have profound implications in real-world systems, such as biological network design or urban network design, which is usually regarded as a generative process.
Particularly, generative AI models such as diffusion models \cite{lee2023score} and auto-regressive approaches~\cite{ingraham2019generative} were proposed to generate novel protein structures.
Meanwhile, network search and design can also be formulated as a decision-making problem, where a novel structure can be synthesized step by step.
For example, Fan \textit{et al}~\cite{fan2023searching} used deep reinforcement learning to find the ground state of complex networks described by the Ising model. Zheng \textit{et al}~\cite{zheng2023spatial,zheng2023road} also used deep reinforcement learning to find optimal community or road structures that can be formulated as networks.

\subsubsection{Future Directions}
Here, we list various prospective directions in which we anticipate AI can yield promising outcomes.
\begin{itemize}
    \item \textbf{Network revival.} As errors and failures are prevalent in real-world systems, it is crucial to revitalize failed networks to recover their fundamental functionalities~\cite{zou2021quenching,sanhedrai2022reviving}, with the goal of strategically intervening in the network with minimal effort to revive a failed network.
    \item \textbf{Dynamics-aware combinatorial optimization.} AI approaches have been widely adopted for solving combinatorial optimization problems~\cite{zhang2023survey,yan2020learning}, however, most of them concentrate on static problems with only the topological structures taken into consideration, overlooking the influence of network dynamics.
\end{itemize}

\section{Application}

This section is divided into four subsections according to different application domains of complex networks, covering both natural systems and human-made systems.

\subsection{Ecological network}

\begin{table}[h]
\small
\setlength\tabcolsep{5pt}
\renewcommand{\arraystretch}{1.00}
\caption{Ecological network dataset statistics including topological information, dynamical information and support applications.}
\vspace{-1em}
\begin{tabular}{ccccc}
\hline
\textbf{Dataset}                  & \textbf{Information}                                  & \textbf{Topology}                     & \textbf{Dynamics} & \textbf{Application}       \\ \hline
{\makecell[c]{Mutualistic \\ network}}            & {\makecell[c]{Interactions among plants and \\ pollinators~(or frugivore, insects)}} & {undirected}            & {None}              & \makecell{Prediction~\cite{saavedra2013estimating,zoller2023plant}\\ Inference~\cite{young2021reconstruction} \\ Control~\cite{jiang2019harnessing}} \\ \hline
\multirow{2}{*}{\makecell[c]{Microbial \\ network}}            & \multirow{2}{*}{\makecell[c]{Species interactions \\ in microbial communities}} & \multirow{2}{*}{\makecell[c]{undirected, \\weighted}}            & \multirow{2}{*}{Realistic}              & \multirow{2}{*}{\makecell[c]{Prediction~\cite{ratzke2020strength}\\ Simulation~\cite{pacheco2019costless}}} \\ 
& & & & \\ \hline
\multirow{2}{*}{\makecell[c]{Food \\ web}}            & \multirow{2}{*}{\makecell[c]{Feeding relationships among \\ species within a community }} & \multirow{2}{*}{directed}            & \multirow{2}{*}{None}              & \multirow{2}{*}{{Prediction~\cite{perkins2022consistent}}} \\ 
& & & & \\ \hline
\multirow{2}{*}{\makecell[c]{Food \\ web stability}}            & \multirow{2}{*}{\makecell[c]{The temporal variation \\of species richness}} & \multirow{2}{*}{directed}            & \multirow{2}{*}{Realistic}              &  \multirow{2}{*}{Prediction~\cite{zhao2023relationships}} \\ 
& & & & \\ \hline
\makecell[c]{Comprehensive \\ ecosystem}            & {\makecell[c]{Predatory, mutualistic, and competitive\\
interactions among species }} & {directed}            & {None}              & Inference~\cite{merz2023disruption} \\ \hline
\end{tabular}
\label{tab:eco}
\vspace{-1em}
\end{table}

\begin{table}[h]
\small
\setlength\tabcolsep{5pt}
\renewcommand{\arraystretch}{1.00}
\caption{Details of applications in ecology networks.}
\vspace{-1em}
\begin{tabular}{cccc}
\hline
\textbf{Paper}                                  & \textbf{Application}               & \textbf{Problem}              & \textbf{Methodology} \\ \hline
\cite{wang2024identifying}		& Keystone species in microbial communities & Prediction                 & Deep learning               \\ \hline
\cite{saavedra2013estimating}  & Environmental change tolerance & Prediction                 & Theoretical framework               \\ \hline
\cite{ratzke2020strength}      & Interaction strength           & Prediction                 & Empirical experiment               \\ \hline
\cite{perkins2022consistent}   & Predator-to-prey biomass       & Prediction                 & Scaling pattern analysis              \\ \hline
\cite{zhao2023relationships}   & Food web stability             & Prediction                 & Linear mixed model               \\ \hline
\cite{zoller2023plant}         & Mutualistic network change     & Prediction                  & Statistical analysis               \\ \hline
\cite{young2021reconstruction} & Mutualistic network structure  & Inference                  & Bayesian statistical inference               \\ \hline
\cite{bischof2020estimating,munch2023constraining}   & Population dynamics            & Simulation                 & \makecell[c]{statistical model fitting, \\metabolic theory-infused ML}               \\ \hline
\cite{rangel2018modeling}      & Biodiversity                   & Simulation                 & mechanistic simulation model               \\ \hline
\cite{pacheco2019costless}     & Interspecies interaction          & Simulation                 & genome-scale metabolic model               \\ \hline
\cite{jiang2019harnessing}     & Tipping points                 & Control & Theoretical framework                \\ \hline
\cite{sanhedrai2022reviving}     & Network revival                 & Control &    Theoretical framework              \\ \hline
\end{tabular}
\label{tab:ecotask}
\vspace{-1em}
\end{table}

Ecological networks describe the complex interplay between species, serving as an important perspective to understand symbiosis, competition, and predation among species in the study of ecosystems. Existing works mainly concentrate on three types of networks: (i) food webs, which are usually directed networks illustrating the predatory relationship between organisms as well as the energy transfer of species in an ecosystem~\cite{perkins2022consistent, o2023warming}.
(ii) mutualistic networks~\cite{zoller2023plant}, which focus on mutualistic relationships between anemone and fish, plants and pollinators/seed dispersers, etc., and facilitate the research of plant reproduction and biodiversity maintenance.
(iii) comprehensive ecosystem networks~\cite{merz2023disruption}, which integrate predatory, mutualistic, and competitive interactions among species, provide a robust framework for understanding the holistic behavior of ecosystems.

Through these empirical networks, extensive works have concentrated on predicting ecosystem responses to environmental changes~\cite{saavedra2013estimating, ratzke2020strength},
guiding conservation efforts~\cite{jiang2019harnessing}, etc. Ecological networks help us understand the resilience and sustainability of ecosystems, highlighting the importance of each species and interaction in maintaining ecosystem health. However, to the best of our knowledge, few works have adopted promising AI techniques to address the challenges in this field. Toward this end, we initially provide a comprehensive summary of representative datasets pertaining to ecological networks~(Table~\ref{tab:eco}), including topological information, dynamical information and support applications. We observe that these datasets have been adopted in various applications that fall into prediction, simulation, inference and control of complex networks. Subsequently, we delve into a detailed analysis of specific applications and emphasize the promising role AI could play~(Table~\ref{tab:ecotask}). Due to space limits, we provide detailed discussion in Appendix~\ref{app:a1}.

\subsection{Biology network}

\begin{table}[h]
\vspace{-1em}
\small
\setlength\tabcolsep{3pt}
\renewcommand{\arraystretch}{1.00}
\caption{Biology network dataset statistics including topological information, dynamical information and support applications.}
\vspace{-1em}
\begin{tabular}{ccccc}
\hline
\textbf{Dataset}                  & \textbf{Information}                                  & \textbf{Topology}                     & \textbf{Dynamics} & \textbf{Application}       \\ \hline
PDB                                  & Protein structure            & undirected & None              & Representation~\cite{pu2019deepdrug3d,ragoza2017protein}  \\ \hline
QM9                                  & Molecular structure          & undirected & None              & Generation~\cite{shi2021learning, luo2022autoregressive}\\ \hline
Drugs                                & Molecular structure          & undirected & None              & Generation~\cite{shi2021learning}                                                                                     \\ \hline
Sequences                            & RNA structure                & undirected & None              & Representation~\cite{hofacker1994fast} \\ \hline
STRING, BioGRID                              &  Protein-to-protein interaction               & undirected & None              & Prediction~\cite{szklarczyk2021string,oughtred2021biogrid} \\ \hline
TRRUST, RegNetwork                               & Gene regulatory             & directed   & None              & Inference~\cite{liu2015regnetwork,han2018trrust}            \\ \hline
\end{tabular}
\label{tab:bio}
\vspace{-1em}
\end{table}

\begin{table}[h]
\small
\setlength\tabcolsep{5pt}
\renewcommand{\arraystretch}{1.00}
\caption{Details of applications in biology networks.}
\vspace{-1em}
\begin{tabular}{cccc}
\hline
\textbf{Paper}                                  & \textbf{Application}               & \textbf{Problem}              & \textbf{Methodology} \\ \hline
\cite{ragoza2017protein, sato2019protein, pu2019deepdrug3d} & Protein structure         & Representation      & 3DCNN                                   \\ \hline
\cite{roohani2023predicting}                                & Transcriptional responses                & Prediction          & GNN                                     \\ \hline
\cite{kamimoto2023dissecting}                               & Cell identity shifts                     & Prediction          & None                                    \\ \hline
\cite{lin2023protein}                                       & Inter-protein residue contacts           & Prediction          & CNN + LM                                \\ \hline
\cite{wang2023deep}                                         & Protein-protein binding                  & Prediciton          & GNN                                     \\ \hline
\cite{li2022inferring, shu2021modeling}                     & Transcription factor regulatory networks & Inference           & GNN, SEM + VAE                          \\ \hline
\cite{ma2023single}                                         & Gene association networks                & Inference           & GNN + Transformer                       \\ \hline
\cite{shi2021learning, luo2022autoregressive}               & 3D molecular conformation                & Generation          & \makecell[c]{Autoregressive model \\Score-based model}  \\ \hline
\end{tabular}
\label{tab:biotask}
\vspace{-1em}
\end{table}

Biology networks illustrate the structures of biomolecules and interactions between them, including DNA/RNAs, micromolecules, and proteins. These entities often have rather complex structures and require activation and mediation through interactions with others. Therefore, understanding their structural properties and interaction networks is key to unraveling the complexities of cellular processes and biological functions. 
Biological networks shed light on the coordination of molecular signaling and regulatory pathways within organisms, which are vital for understanding disease mechanisms, advancing drug discovery, and formulating targeted therapies. 
Furthermore, they offer perspectives on the evolutionary connections and overlapping functions of biomolecules, thereby deepening our understanding of biological diversity and adaptability. 
Existing works mainly concentrate on two meta-types of biological networks: (i) structural networks, which describe the structure of proteins, RNAs, and micromolecules; (ii) interaction networks, which demonstrate the complex protein-protein, protein-RNA interactions.

Existing works have focused on AI techniques, especially deep learning on graphs, to address problems in the realms of protein structure prediction~\cite{ wang2023deep}, network inference~\cite{li2022inferring, shu2021modeling, ma2023single}, and drug design~\cite{shi2021learning, luo2022autoregressive}. Collectively, these studies underscore the formidable capabilities and potential of AI methodologies in advancing scientific understanding and innovation in these areas. We provide a comprehensive summary of representative datasets pertaining to biology networks~(Table~\ref{tab:bio}), including topological information, dynamical information and support applications. We observe that these datasets have been adopted in various applications that fall into representation, prediction, inference and generation of complex networks. The details of these specific applications regarding the utilized AI models are listed in Table~\ref{tab:biotask}. Due to space limits, we provide detailed discussion in Appendix~\ref{app:a2}.

\subsection{Urban network}

\begin{table}[h]
\small
\setlength\tabcolsep{1pt}
\renewcommand{\arraystretch}{1.1}
\caption{Urban network dataset statistics including topological information, dynamical information and support applications.}
\vspace{-1em}
\begin{tabular}{ccccc}
\hline
\textbf{Dataset}                  & \textbf{Information}                                  & \textbf{Topology}                     & \textbf{Dynamics} & \textbf{Application}       \\ \hline
PyPSA-Eur~\cite{horsch2018pypsa} & \makecell[c]{The European electric \\ transmission system}     & {\makecell[c]{Directed, \\weighted}}  & Realistic            & {\makecell[c]{Prediction~\cite{silva2021combining}, \\ Generation~\cite{kocayusufoglu2022flowgen}}} \\ \hline
PowerGraph~\cite{amara2023powergraph}    & \makecell[c]{Power grid benchmark \\ dataset for GNN}     & {\makecell[c]{Directed, \\weighted}}      & Realistic    & Prediction\cite{varbella2023geometric}  \\ \hline
Road traffic~\cite{li2023dynamic,li2017diffusion}                                  & \makecell[c]{Road traffic dataset in \\ Beijing, L.A. and Bay area}            & {\makecell[c]{Directed, \\weighted}} & Realistic & {\makecell[c]{Prediction~\cite{wu2021inductive,jin2023dual,chen2020autoreservoir}, \\ Simulation~\cite{chen2020autoreservoir}}}    \\ \hline
GenNet~\cite{li2023gennet} & \makecell[c]{Synthetic data ecosystem \\ for mobile networks} & {\makecell[c]{Directed, \\dynamic}} & Realistic                &  \makecell[c]{Generation~\cite{long2023practical,hui2023large,zhang2023deep}, \\ Decision-making~\cite{huang2023safe}} \\ \hline
GUI~\cite{han2023GUI} & \makecell[c]{A comprehensive dataset of \\ global urban infrastructure } & Directed & None               & Decision-making~\cite{mao2023detecting}  \\ \hline
\makecell[c]{COVID19-mobility-US\\\cite{kang2020multiscale}  }                                & \makecell[c]{Multiscale human mobility \\
flow in the US}            & {\makecell[c]{Directed, \\dynamic}} & Realistic              & Generation~\cite{simini2021deep} \\ \hline
 \makecell[c]{Urban-mobility-US\\\cite{safegraph2023social,chang2021mobility,rong2023interdisciplinary}} & \makecell[c]{Urban mobility data collected \\ by SafeGraph, Cuebiq and \\ government census \\ in the US}            & {\makecell[c]{Directed, \\dynamic \\ or static}} & Realistic               & \makecell[c]{Prediction~\cite{shah2020finding} \\ Simulation~\cite{huang2021coupled} \\ Generation~\cite{rong2023complexity}  \\  Decision-making~\cite{hao2023gat,hao2022reinforcement}} \\ \hline
 \makecell[c]{Urban-mobility-Spain\\\cite{spain2023covid}} & \makecell[c]{Urban mobility flow \\ in the Spain}             & {\makecell[c]{Directed, \\dynamic}} & Realistic             & Simulation~\cite{murphy2021deep,heredia2023forecasting}\\ \hline
\end{tabular}
\label{tab:dataset_unet}
\vspace{-1em}
\end{table}

\begin{table}[h]
\small
\setlength\tabcolsep{5pt}
\renewcommand{\arraystretch}{1.00}
\caption{Details of applications in urban networks.}
\vspace{-1em}
\begin{tabular}{cccc}
\hline
\textbf{Paper}                                  & \textbf{Application}               & \textbf{Problem}              & \textbf{Methodology} \\ \hline
\cite{varbella2023geometric}                        &  Power grid cascading failure                & Prediction         & GCN        \\ \hline
\cite{bassamzadeh2017multiscale}                     & Electricity demand                & Prediction         & Bayesian network                                     \\ \hline
\cite{jin2023dual}  & Transportation travel time & Prediction & GCN\\ \hline
\cite{wu2021inductive}      &Spatio-temporal kriging        &Prediction     & GCN \\ \hline
\cite{shah2020finding} & Epidemic source identification  &  Prediction & GCN \\ \hline
\cite{silva2021combining}  & Electric network flow &  Prediction          & GCN                                     \\ \hline
\cite{chen2024social} & Crowd movements         & Simulation      & Diffusion model, Equivariant GNN                                   \\ \hline
\cite{chen2020autoreservoir} & Wind speed, traffic flow         & Simulation      &  Auto-reservoir neural network                                   \\ \hline
\cite{murphy2021deep,fu2023privacy} & Contagion dynamics         & Simulation      & GCN                                   \\ \hline
\cite{huang2021coupled} & Contagion dynamics         & Simulation      & ODE + GCN                                   \\ \hline
\cite{kocayusufoglu2022flowgen} & Electric network flow &  Generation          & GAN                                     \\ \hline
\cite{simini2021deep,rong2023complexity}           & Urban mobility flow  & Generation  & MLP, GCN, Diffusion model \\ \hline
\cite{mao2023detecting}                     & Vulnerable node detection                & Decision-making         & GCN + RL                                      \\ \hline
\cite{hao2023gat,hao2022reinforcement} &  Covid-19 vaccine allocation & Decision-making  & GCN + RL  \\ \hline
\cite{li2023carbon} & Network energy efficiency optimization & Decision-making  & GCN + RL  \\ \hline
\end{tabular}
\label{tab:urbantask}
\vspace{-2em}
\end{table}

Numerous elements in the city take the form of networks, primarily divided into two categories: networks formed by various urban infrastructures and networks shaped by human activities within the city. Urban infrastructure networks include critical engineering facilities for the normal operation of a city, such as transportation, power supply, communication, water supply, and drainage. Human activities in the city contribute to forming another large category of urban networks, such as the mobility flow network between urban regions. More specifically, human commuting behaviors result in the formation of a commuting flow between regions, human movements during disasters result in the formation of an evacuation flow between regions, human taxi and cycling behaviors result in the formation of original-destination~(OD) flow between regions, and encounters between humans also lead to the formation of human contact networks.
Moreover, diverse urban infrastructure networks and urban activity networks are tightly coupled with each other. Altogether, these coupled urban networks compose a typical interdependent network. Consequently, the functional failure of a single facility under the impact of disasters or accidents often not only influences local regions but also affects a large area of the city. Thus, investigating urban networks is of great importance.

The summary of representative datasets of urban networks is provided in Table~\ref{tab:dataset_unet}, including topological information, dynamical information and support applications that fall into the prediction, simulation, generation and decision-making of complex networks. The details of these specific applications regarding the utilized AI models are listed in Table~\ref{tab:urbantask}. Due to space limits, we provide detailed discussion in Appendix~\ref{app:a3}.

\subsection{Society network}

\begin{table}[h]
\small
\setlength\tabcolsep{1pt}
\renewcommand{\arraystretch}{1.1}
\caption{Society network dataset statistics including topological information, dynamical information and support applications.}
\vspace{-1em}
\begin{tabular}{ccccc}
\hline
\textbf{Dataset}                  & \textbf{Information}                                  & \textbf{Topology}                     & \textbf{Dynamics} & \textbf{Application}       \\ \hline
\makecell[c]{Online social network\\\cite{xie2021detecting,flamino2023political,bovet2019influence}} & Twitter and Weibo  &   {\makecell[c]{Directed, \\weighted}}  &  Realistic & Prediction~\cite{wang2022casseqgcn}  \\ \hline
Social e-commerce~\cite{xu2019relation} &  \makecell[c]{A e-commerce platform, Beidian \\ based on social recommendation}   &   {\makecell[c]{Directed, \\weighted}}  &  Realistic & Prediction~\cite{xu2019relation} \\ \hline
FakeNewsNet~\cite{shu2020fakenewsnet} & \makecell[c]{content, social context, and \\spatiotemporal information of fake news} & {\makecell[c]{Directed, \\weighted}} & None & Prediction~\cite{wu2023decor} \\ \hline
\makecell[c]{Supply chain\\\cite{xu2019resiliency,zhang2022estimating,li2023learning}}  & \makecell[c]{Relationship between producers \\along supply chains}  &  Directed   &   Simulated & Prediction~\cite{li2023learning} \\ \hline
\makecell[c]{Scientific collaboration\\~\cite{mag,priem2022openalex,web-of-science}} &   \makecell[c]{Microsoft Academic Graph, \\OpenAlex, and Web of Science}   & {\makecell[c]{Directed, \\weighted}}   & None  & Prediction~\cite{zhang2019oag} \\ \hline
\end{tabular}
\label{tab:soc}
\vspace{-1em}
\end{table}

\begin{table}[h]
\small
\setlength\tabcolsep{8pt}
\renewcommand{\arraystretch}{1.1}
\caption{Details of applications in society networks.}
\vspace{-1em}
\begin{tabular}{cccc}
\hline
\textbf{Paper}                                  & \textbf{Application}               & \textbf{Problem}              & \textbf{Methodology} \\ \hline
\cite{yang2021full}  & Information propagation & Prediction & RNN + RL \\ \hline
\cite{ji2023community} & Information propagation & Prediction & GNN \\ \hline
\cite{zhang2023attentional} & Friendship relation & Prediction & GNN \\ \hline
\cite{wu2023decor} & False news detection & Prediction & GNN \\ \hline
\cite{li2023learning} & Supply chain & Prediction & Temporal KG embedding, RNN \\ \hline
\cite{okawa2022predicting} & Opinion dynamics & Simulation & Neural ODE \\ \hline
\cite{park2023generative} & Individual social behavior & Simulation & LLM \\ \hline
\cite{gao2023s} & Collective social behavior & Simulation & LLM \\ \hline
\cite{fan2020finding} & Network dismantling & Decision-making & GNN + RL \\ \hline 
\cite{chen2021contingency,ling2023deep} & Information maximization & Control & GNN + RL  \\ \hline
\end{tabular}
\label{tab:soctasks}
\vspace{-1em}
\end{table}

Society networks, characterized by complex social interactions among individuals or organizations, have been the focus of research on complex networks for decades. Prior work mainly focuses on three types of society networks: (i) social networks, where individuals interact, share information and engage in social activities both in person and through digital means, (ii) scientific collaboration networks, where scientists, researchers, and organizations collaborate together to advance scientific discovery, foster innovation, and accelerate the pace of research, (iii) global economic networks, where firms, countries, and regions are connected through financial and trade relationships in order to facilitate the flow of goods, services, capital, and information. Though nodes in these networks vary from individuals to organizations and even to countries, they share many common properties and underlying mechanisms. Therefore, we group them together for further discussion.

The recent development of AI has proven its great potential to handle challenges, including lack of exact solutions, unknown dynamics and noisy data. Therefore, incorporating AI into the modeling of complex society networks is a promising direction.
The summary of representative datasets of society networks is provided in Table~\ref{tab:soc}, including topological information, dynamical information and support applications that mainly focus on prediction of complex networks. The details of these specific applications regarding the utilized AI models are listed in Table~\ref{tab:soctasks}. Due to space limits, we provide detailed discussion in Appendix~\ref{app:a4}.

\section{Discussion}

\subsection{Combining with mechanistic approach}

Existing methods such as symbolic regression~\cite{gao2022autonomous} have demonstrated that we can leverage AI techniques to discover unknown mechanistic models in complex networks. Consequently, it is feasible to gradually replace components in a model derived from pure neural networks with thoroughly analyzed mechanistic models. By integrating these mechanistic models and examining the derived properties, we can validate the correctness of the discovered mechanistic models and further calibrate them. By repeating the steps of discovery and calibration, we can finally gain a thorough understanding of the complex networks.

Another possibility is to use different mechanistic models for different parts of a complex network, such as macroscale and microscale. In scenarios where a known mechanistic model exists at a specific scale but the mechanisms at other scales are unclear, the physics-informed machine learning paradigm can effectively integrate underlying mechanisms with machine learning methods, enhancing the modeling capabilities of complex networks. Recent works have pointed out that high-dimensional complex systems  are intrinsically low-rank~\cite{thibeault2024low,lyu2024learning}, suggesting a new possibility of hybrid approaches.

Simultaneously, combining AI technologies with mechanistic approaches enhances their interpretability and generalization capabilities significantly. For example, in Hamiltonian neural networks~\cite{greydanus2019hamiltonian}, Hamiltonian mechanics is integrated into neural networks, effectively preserving conservation laws in the complex system. It is also a promising direction for analyzing complex networks using AI methods. By integrating classical theories and mechanistic models of complex networks, e.g., scale-free, self-similarity, with AI methods, we can yield an approach that achieves interpretability, generalization capabilities, and modeling accuracy.

\subsection{Complex network for AI}

As more and more AI technologies are being applied, they themselves have formed into complex systems, requiring a complexity science perspective to understand AI and address the issues in its development. Here, we summarize three potential research directions, covering structural analysis of complex AI models, understanding of optimization process, and recent highly regarded research on LLMs.

First, based on established theories regarding complex network topology, it is possible to \textit{investigate whether well-performing deep learning networks exhibit universal structural patterns}. Sun et al.~\cite{sun2020revealing} used information theory and entropy-related mechanisms from statistical physics to study the limits of network structure prediction, deriving performance bounds that can be used to guide the design of machine learning algorithms in scenarios like recommendation systems and protein-protein interaction detection. The connectivity of neural networks is also an important aspect affecting their performance~\cite{you2020graph,scabini2023structure}, inspired by the connectionist AI ideas influenced by biological neural networks. In the field of AI, researchers have observed that randomly wiring topology~\cite{xie2019exploring} or sparse connectivity~\cite{mocanu2018scalable} inspired by network science can surprisingly induce empirically well-performed neural networks. A recent study on the feedforward connectivity map of the mouse cerebellar cortex has observed a redundant, non-random connectivity pattern, which increases noise resilience without significantly reducing network encoding capacity~\cite{nguyen2023structured}. The balance between encoding capacity and redundancy can guide the design of artificial neural networks, helping us answer key questions about the mechanisms and network connectivity patterns that determine AI intelligence levels. In terms of interpreting black-box AI models,  utilizing topological analysis in high-dimensional space can aid the understanding of complex correlations behind model prediction results~\cite{liu2023topological}.

Second, from the system dynamics perspective, statistical physics theories highly related to complex systems can be used to \textit{analyze the optimization process of AI models}. Two early representatives in this regard are the Hebbian learning rule~\cite{hebb1949organization} and the Hopfield network \cite{hopfield1982neural}, which fall into rather idealized approaches. Recent work has used a scale separation approach of fast and slow variables to explain and predict the feature learning ability of deep convolutional networks using thermodynamics theory~\cite{seroussi2023separation}. The updating of the parameter weights in neural networks can be viewed as a dynamic evolutionary process. Yang et al.~\cite{yang2023stochastic} studied the underlying learning dynamics of the stochastic gradient descent~(SGD) algorithm, and identified the source of SGD's favor in flat solutions by solving the Fokker-Planck equation. Durr et al.~\cite{durr2023effective} analyzed the learning dynamics of adversarial training in GANs and identified the conditions of mode collapse. Therefore, it is possible to study whether there are regularities from a dynamical systems perspective, which can be used to inspire new methods for designing and training AI models.

Finally, \textit{exploring the relationship between LLMs and complex networks} will be an important research frontier. On one hand, LLMs have the potential to become important research tools in complexity science. Although existing LLMs cannot directly answer complex science questions, they have the powerful capability of collaborating and interacting with humans, storing vast knowledge, and retrieving it accurately. Therefore, they can inspire scientists to propose new complex systems theories, as in the recent advance of LLM-based mathematical discovery of cap set problem~\cite{romera2023mathematical}. Recent studies have used chatGPT to build a society of AI agents and simulate corresponding collective behaviors~\cite{gao2023large}, indicating the great potential of LLMs in understanding and developing complexity science.
On the other hand, research on complex networks can also contribute to developing LLMs. Current research on the emergent capability of LLMs falls into empirical data observation and analysis~\cite{wei2022emergent}. The underlying reasons may correspond to the information flow efficiency and structural characteristics of neural networks. Thus, it is possible to consider conducting research using the mindset and tools of complexity science.

\subsection{Network science discovery in the age of AI}
In the realm of network science, AI has been proven useful in simulating temporal evolution~\cite{murphy2021deep}, making inferences with unknown network dynamics~\cite{gao2022autonomous}, and solving high-dimensional combinatorial optimization problems~\cite{fan2020finding}. However, generating scientific hypotheses is crucial to scientific discovery, which is missing in the above works. Comparatively, they can only combine several existing formulas in a manually designed library to generate a new form of dynamic function or search for improved solutions in a high-dimensional space based on a black-box AI method.
Therefore, it remains an unresolved challenge to discover new theories of network science. In this regard, utilizing generative AI technology or LLMs can be a promising solution as they can generate both effective and human-readable hypotheses for human experts.

\section{Conclusion}
This survey systematically discusses the potential advantages of AI in tackling unresolved challenges of complex network research, summarizes the key research problems of AI for complex networks, and provides a comprehensive review of corresponding methodologies and applications. We expect the transformative capability of AI will play a more vital role in network science.

\bibliographystyle{plain} 
\bibliography{ref}
\appendix
\counterwithin*{table}{section} 
\renewcommand{\thetable}{S\arabic{table}} 
\section{Related survey}
\begin{table}[h]
\small
\setlength\tabcolsep{5pt}
\renewcommand{\arraystretch}{1.05}
\caption{A list of related surveys.}
\begin{tabular}{ccc}
\toprule
\textbf{Domain} & \textbf{Category} & \textbf{Survey} \\
\toprule
\multirow{9}{*}{\makecell[c]{Complex \\ network}} & Comprehensive & \makecell[l]{statistical mechanics~\cite{albert2002statistical}, structure and dynamics~\cite{boccaletti2006complex},\\ data mining in complex network~\cite{zanin2016combining}} \\ \cline{2-3} 
& Theory & \makecell[l]{coevolving network~\cite{maslennikov2017adaptive}, high-order interactions~\cite{battiston2020networks,battiston2021physics,boccaletti2023structure},\\ coupled dynamical network~\cite{zou2021quenching}, representation~\cite{comin2020complex,torres2021and},\\ message passing~\cite{newman2023message}, robustness and resilience~\cite{artime2024robustness}} \\ \cline{2-3} 
& Application & \makecell[l]{coevolution spreading~\cite{wang2019coevolution}, network reconstruction~\cite{squartini2018reconstruction},\\ vital node itentification~\cite{lu2016vital}, dynamic network recovery~\cite{wu2021recovering},\\ link prediction~\cite{lu2011link}, structure prediction~\cite{ren2018structure}, control network~\cite{d2023controlling},\\ dynamics inference~\cite{gao2023data}, signal propagation~\cite{ji2023signal}, hypergraph mining~\cite{lee2024survey}}\\ \midrule

\multirow{3}{*}{AI} & Methodology & \makecell[l]{graph embedding~\cite{cai2018comprehensive}, GNN~\cite{wu2020comprehensive}, RL on graphs~\cite{wu2020comprehensive},\\ generative model~\cite{guo2022systematic}, physics-informed machine learning~\cite{karniadakis2021physics}} \\ \cline{2-3} 
& AI for science & scientific discovery~\cite{xu2021artificial,wang2023scientific}, ecology~\cite{han2023synergistic}  \\ \midrule

\end{tabular}%
\label{survey}
\end{table}

\section{Methodology details}
\subsection{Representation problem} \label{app:m1}
\textbf{Simplicial complex.}
A Simplicial complex can be seen as a special case of hypergraphs, which requires the hyperedge set $\mathcal{H}$ to be closed under inclusion. Specifically, if $h\in\mathcal{H}$ is a hyperedge, then any non-empty subset should also be a hyperedge:
\begin{equation}
    h\in\mathcal{H} \Rightarrow \forall h'\subseteq h, h'\neq\emptyset, h'\in\mathcal{H}
\end{equation}
For example, if $h=[a, b, c]\in\mathcal{H}$, then $[a], [b], [c], [a,b], [b,c], [a,c]$ belongs to $\mathcal{H}$ as well.
This stronger assumption restricts the potential application scenarios of simplicial complexes. However, it also equips researchers with powerful mathematical tools for learning representations on simplicial complexes.
For example, Piaggesi et al.\cite{piaggesi2022effective} extended the Word2Vec\cite{mikolov2013distributed} and DeepWalk\cite{perozzi2014deepwalk} methods. They proposed a random walk-based approach by leveraging the Hasse Diagram, which essentially gives an extension of the concept of adjacency matrices in the realm of simplicial complexes.
In contrast, Zhou et al.\cite{zhou2023facilitating} harnessed the Hodge Laplacian, which naturally extends the Graph Laplacian ($L=D-A$, where $D$ is the degree matrix and $A$ is the adjacency matrix of a graph) according to discrete Hodge-deRham theory. They used Hodge Laplacian to extend existing positional encoding (PE) and structure encoding (SE) methods into high-order ones and thus facilitate GNN-based representation learning.

\textbf{Subgraph.}
Rossi et al.\cite{rossi2020structural} proposed to learn node representation with motif matrix, which can be formally defined as
\begin{equation}
    W_{ij}^k = \sum_{m \in M_k} \mathbb{I}(v_i \in m, v_j \in m),\quad \forall i, j \in \{1, \cdots, |\mathcal{V}|\}, k\in{1, \dots, K},
\end{equation}
where $M_k$ is the set of all motifs of type $k$. In other words, $W_{ij}^{k}=n$ indicates that nodes $v_i$ and $v_j$ participates in the $k$-th motif type $n$ times. They solved the low-rank decomposition of all motif matrices and concatenated them together to get the node representation that preserves the structural similarity among nodes.
Liu et al.\cite{liu2021motif} used motif matrix as well, but in contrast, they added the motif matrixes $W^k$ of all motif types and the adjacency matrix $A$ together to get a trainable linear transformation weight matrix
\begin{equation}
    W = A + \beta \sum_{k=1}^K \alpha_k W^k, \quad (\alpha_k, \beta \mathrm{\ are \ trainable \ parameters}),
\end{equation}
and adopted it as the message-passing matrix in GCN to learn the node representation that preserves the local high-order structures.
Cotta et al.\cite{cotta2020unsupervised} proposed a GNN-based method to learn representations for motifs to get joint representations of $k>2$ nodes. Specifically, for a given network, they generated a new network called $k$-HON, whose vertices correspond to subgraphs with $k$ nodes in the raw network, and edges represent the neighborhood relationships between these subgraphs. They then use GNNs to learn the node representation of the $k$-HON network and thus get the joint representations of $k$ nodes, i.e. the representations of motifs,  in the raw network.
On the other hand, Yu et al.\cite{yu2022molecular} focused on learning representations for chemical molecules, while still adopting a GNN-based method to learn representations for motifs. They defined motifs as some typical subgraphs in molecules that have specific chemical functions, and generate a heterogeneous graph that contains both the molecules and the motifs. They used a GNN to learn the representation of motifs and molecules simultaneously and found that considering motifs can improve the performance of downstream tasks such as molecular property prediction.

\textbf{Temporal variation}
Liu et al.\cite{liu2021motif} used GNN to learn node representation for a dynamic graph, and introduced a Temporal Shift mechanism to model the temporal dynamics. Specifically, they use GNNs to learn node representation for each time snapshot, when a GNN conducts message passing, it can leverage not only the latent embedding at the current time snapshot but also the previous time snapshot.
Wen and Fang\cite{wen2022trend} also enabled GNNs to leverage the latent embedding of historical time snapshots as well as the current time snapshots to learn node representation for a dynamic graph. But they used not only the previous time snapshot but all historical time snapshots, and leveraged a kernel function $\kappa(\Delta t) = \exp(-\delta \Delta t)$ to weight the historical time snapshots according to the time interval $\Delta t$ between the current time snapshot and the historical time snapshot, where $\delta$ is a trainable parameter.
Instead of capturing the relationship between node representations in different time snapshots with neural networks directly, Li et al.\cite{li2023scaling} designed an embedding updating process inspired by the integrate-and-fire model. Specifically, given the representation $h_{i}^{t}$ of node $v_i$ at time $t$, they aggregate the message $m_{ji}^t$ passed from its neighbours $v_j$ and update the representation of $v_i$ by
\begin{equation}
    \tau_m(h_{i}^{t+1}-h_{i}^{t}) = -(h_{i}^{t} - V_\mathrm{reset}) + \mathrm{GNN}(h_i^t, \sum_{j\in\mathcal{N}(i)} m_{ji}^t)
\end{equation}
where $\tau_m$ and $V_\mathrm{reset}$ are two hyperparameters. If $h_i^{t+1}$ is greater than a threshold $V_\mathrm{th}$, set $h_i^{t+1}$ to $V_\mathrm{reset}$ and generate a boolean value $m_{ik}^{t+1}=1$ as the message passed to all of its neighbours $v_k$ at next time snapshot. Otherwise, set $m_{ik}^{t+1}=0$. They found that this model can significantly decrease parameters and computation overheads and generalize to a large temporal graph.

\subsection{Prediction problem} \label{app:m2}
\textbf{Node state prediction.}
Dong et al.~\cite{dong2019multiple} modeled source localization as a node classification problem with node features $y_T$ and node classes $x$ and adopted GCNs to solve it. Apart from $y_T$, they also leverage domain knowledge (label propagation-based source identification) to design a new node feature $(1-\alpha)(I-\alpha L)y_T$ that captures the feature of centrality, where $\alpha$ is a hyperparameter and $L$ is the normalized Laplacian matrix of the network.
Instead of localizing sources directly, other works proposed to address it as an inverse problem of graph diffusion. 
For example, Wang et al.\cite{wang2022invertible} introduced a graph diffusion model
\begin{equation}
    y_T = \frac{1}{2}(\zeta+\mathrm{GNN}(\zeta; W_2)),\quad \zeta = \frac{1}{2}(x+\mathrm{GNN}(x; W_1)),
\end{equation}
where $W_{1, 2}$ are trainable parameters. After trained, the diffusion model can be inversed with two fix point iterations to get a source node estimation $x^{M}$ from the diffusion vector $y_T$:
\begin{equation}
    \zeta^{i} = 2y_T - \mathrm{GNN}(\zeta^{i-1}; W_2), \quad x^{i} = 2\zeta - \mathrm{GNN}(x^{i-1}; W_1), \quad \text{repeat for $M$ times ($i \in {1..M}$)}.
\end{equation}
Huang et al.\cite{huang2023two} integrated forward (source localization) and inverse (graph diffusion) perspectives together and proposed a two-stage approach: In the coarse stage, a GNN model generates coarse-grained source predictions, while in the fine stage, a generative diffusion model quantifies the prediction uncertainty by simulating the dissemination process to get a finer result.
For critical node detection, the definition of critical nodes in different fields also varies correspondingly.
Munikoti et al.\cite{munikoti2022scalable} defined critical nodes as those whose removal substantially decreases network robustness~(measured by effective graph resistance and weighted spectrum) and employed GNNs to learn the criticality score of nodes in an end-to-end fashion.
Zhao et al.\cite{zhao2023novel} adopted a similar definition, but they used the motif matrix rather than the adjacency matrix for message-passing to consider high-order interactions. 
Regarding social influence, Qiu et al.\cite{qiu2018deepinf} defined critical nodes as those that can be easily influenced by their neighbors. 
Specifically, they break the network $G=(\mathcal{V}, \mathcal{E})$ into ego-networks $G_i = (\mathcal{V}_i, \mathcal{E}_i)$ for each node $v_i$, where $\mathcal{V}_i$ is the set of nodes that can reach $v_i$ within $k$ hops and $\mathcal{E}_i = (\mathcal{V}_i \times \mathcal{V}_i) \cap \mathcal{E}$, and proposed a GCN-based end-to-end framework to estimate that given the ego-network $G_i$ and the neighbour nodes' activation states $s_j^{t}\in\{0,1\}~(j \in \mathcal{V}_i)$ at time $t$, whether $v_i$ will be activated before time $t+\Delta t$. Formally, they trained the model with the following loss function:
\begin{equation}
    \mathrm{minimize}_\Theta -\sum_{i} \log P_\Theta(s_v^{t+\Delta t}|G_i, \{s_j^{t}, j\in\mathcal{V}_i\}).
\end{equation}

\textbf{Link formation prediction.}
Some work leveraged graph embedding methods to predict missing links. For example,
Hamilton et al.\cite{hamilton2017inductive} proposed an inductive framework called GraphSAGE (SAmple and aggreGatE) that learns node embeddings by sampling a fixed-size set of nodes' neighbors and aggregating their embeddings.
In contrast, Qiu et al.\cite{qiu2018network} a NetMF method which generates embeddings by matrix factorization. They proved that typical random walk-based graph embedding methods such as DeepWalk\cite{perozzi2014deepwalk} and node2vec\cite{grover2016node2vec} can be actually viewed as implicit factorization of a closed-form matrix when the length of random walks goes to infinity, and further proposed to factorize the closed-form matrix explicitly to get the node embedding with a better performance.
However, none of these methods are efficient enough to scale to large networks with tens of millions of nodes. To address this problem, Qiu et al. further\cite{qiu2019netsmf} proposed a NetSMF method, which specifies the dense matrix used by NetMF while keeping spectrally close to the original one to reduce the computational complexity. NetSMF can largely reduce the computational complexity and scale to large networks while maintaining the link prediction performance compared to NetMF.
Apart from these graph embedding-based methods, there are also works predicting missing links with GNN in a supervised way. In these works, to predict whether two nodes $(u, v)$ in a network $(\mathcal{V}, \mathcal{E})$ are likely to have a link, researchers typically consider the local subgraph of their $h$-hop neighbors $(\mathcal{V}_{u,v}^h, \mathcal{E}_{u, v}^h)$:
\begin{equation}
    \mathcal{V}_{u, v}^h = \{u, v\} \cup \mathcal{N}_{u}^h \cup \mathcal{N}_{v}^h, \quad \mathcal{E}_{u, v}^h = (\mathcal{V}_{u, v}^h \times \mathcal{V}_{u, v}^h) \cap \mathcal{E},
\end{equation}
where $\mathcal{N}_{u}^h$ is the set of nodes that can reach $u$ within $h$ hops. For example,
to predict the link between nodes $(u, v)$, Zhang et al.\cite{zhang2017Weisfeiler} proposed to extract the subgraph induced by their $h$-hop neighbors and then apply an MLP to the adjacency matrix of this subgraph to predict the likelihood of the link between them.
Zhang and Chen\cite{zhang2018link} replaced the MLP with GNNs to empower it to handle local subgraphs with different sizes and learn from node features as well as an adjacency matrix. They also theoretically proved that remote parts of the network contribute little to link existence, and thus a small $h$ is enough to make accurate predictions.
Wang et al.\cite{wang2023neural} proposed a Neo-GNN method inspired by the classical Common Neighbour (CN) method. CN predicts the likelihood of the link between two nodes with a score function that counts their common neighbors, i.e.
\begin{equation}
    S_\mathrm{CN}(i, j) = \sum_{u \in \mathcal{N}_i^1 \cap \mathcal{N}_j^1} 1.
\end{equation}
In contrast, Neo-GNN considers multi-hop common neighbors and introduces a learnable function to improve performance:
\begin{equation}
    S_\mathrm{Neo\mbox{-}GNN}(i, j) = \sum_{l_1=1}^l \sum_{l_2=1}^l \beta^{l_1+l_2-2} z_{l_1, l_2}(i, j), \quad z_{l_1, l_2}(i, j) = \sum_{u \in \mathcal{N}_i^{l_1} \cap \mathcal{N}_j^{l_2}} A^{l_1}_{iu} A^{l_2}_{ju} f(d_u),
\end{equation}
where $A$ is the adjacency matrix, $d_u$ is the degree of node $u$, $f$ is a learnable function, $\beta$ and $l$ are hyperparameters.

\textbf{Network property prediction.}
Ahmad et al.\cite{ahmad2021predicting} introduced a spatio-temporal GCN to predict the potential cascading failure in power systems. 
Specifically, they first constructed a graph to represent the power system, where each node is a bus with a specific load and/or generator and each edge is a transmission line or a transformer. They then use a GCN to extract the spatial features of each node and an LSTM to incorporate temporal information. The output is finally fed into a classifier to predict whether a cascade failure will happen.
Liu et al.\cite{liu2021searching} also focused on the cascade failure in power systems while they aimed to identify as many critical cascading failures as possible for a given network. 
To achieve this goal, they proposed a GNN-based framework that can learn the vulnerability of nodes, and further leverage the vulnerability to heuristically search for the critical cascading failures.
Panagopoulos et al.\cite{panagopoulos2021transfer} proposed a GNN-based model to predict the outbreak of COVID-19 in different countries. Specifically, they first constructed a graph where nodes correspond to regions and edge weights to mass mobility between their endpoints and then adopted a family of GNNs to learn the node representation based on their connectivity and history. Based on these representations, they can predict whether COVID-19 will outbreak in the graph with another classifier. They also leveraged Model-Agnostic Meta-Learning (MAML) to make their model generalizable among different countries, which addressed the problem that the availability of data is limited at the start of the outbreak in a country.
Wang et al.\cite{wang2022causalgnn} focused on a similar problem and adopted a GNN-based model as well. However, they further considered the underlying causal mechanisms to solve the challenge that epidemic data is sparse and noisy and a deep learning model thus tends to overfit. Specifically, they leveraged a causal model such as susceptible(S)-infected(I)-recovered(R)-deceased(D) (SIRD) or SIR to predict the infected cases and leveraged a GNN to generate the residual between the predicted and observed infected cases.

\subsection{Simulation problem} \label{app:m3}
\textbf{Temporal factor.}
LG-ODE~\cite{huang2020learning} designs a temporal self-attention module to encode the temporal pattern of each observation sequence and employs a Graph Neural ODE to model the continuous dynamics in the hidden space. It connects the observed nodes as temporal graphs according to the given topology within a defined time window, thus capturing the temporal dependence between the nodes.

\textbf{Topology factor.}
For noisy information that exists in observed topology, many methods guide the optimization of the topology with the accuracy of the downstream task. TO-GCN~\cite{yang2019topology} refines the given network topology for optimal performance while updating the GCN parameters in the classification task. Similarly, rGNN~\cite{peng2022reverse} learns a new graph that maintains semantic consistency and structural consistency through node features and smoothly adds it to the initial graph to repair misconnections.

\subsection{Inference problem} \label{app:m4}
\textbf{Topology Inference.}
For \textbf{similarity metric} based approaches, the distance of low-dimensional node representations in Euclidean space as a metric can improve computational efficiency when the network size is large~\cite{cosmo2020latent}. Furthermore, distances on product manifolds of constant curvature model spaces are a more efficient similarity metric than those on Euclidean spaces~\cite{de2022latent}. Instead of choosing a single similarity metric, various similarity metrics are fused in Grale~\cite{halcrow2020grale}. To ensure sparsity, one may construct a KNN similarity graph or connect only pairs of nodes that exceed the similarity threshold~\cite{wu2018quest, halcrow2020grale}.
\textbf{Domain knowledge} is leveraged to guide topology inference in some specific applications~\cite{fatemi2021slaps}. For example, in ProGNN~\cite{jin2020graph}, low-rank and sparse properties are applied to infer clean graph structures from poisonous topologies, and syntax trees and regular language can guide graph structure inference for programming programs~\cite{johnson2020learning}.

\subsection{Generation problem} \label{app:m5}
\textbf{Static undirected network.}
For deep generative modeling approaches, 
You et al.~\cite{you2018graphrnn} propose a deep autoregressive model GraphRNN, which decomposes the network generation process into a sequence of node and edge formations, each step being conditioned on the network structure that has been generated up to that point. Liao et al.~\cite{liao2019efficient} developed GRAN based on a recurrent neural network and attention mechanism to generate networks block by block, each containing a group of nodes and their connected edges. This approach allows for a balance between sample quality and efficiency through adjustable block size and sampling stride. Martinkus et al.~\cite{martinkus2022spectre} employ a GAN-based model to tackle the network generation problem from a spectral viewpoint. Their approach starts with generating the dominant components of the graph Laplacian spectrum and then constructs a graph that matches these eigenvalues and eigenvectors.

Beyond the generic structure that only considers the interplay between nodes and edges, networks with attributes correspond to various real-world entities such as proteins, molecules, transportation systems, and ecological systems, illustrating their broad relevance to the real world. Consequently, attributed network generation has attracted significant attention from researchers.
Vignac et al.~\cite{vignac2022digress} present DiGress, a discrete denoising diffusion model for generating networks with categorical node and edge attributes via progressively editing graphs with noisy structures through a process that involves adding or removing edges and modifying their categories. Experimental results indicate that the model can not only learn to generate structures such as grid or planar networks but also shows promising potential for application in real-world molecular design.
Furthermore, recent work~\cite{DBLP:conf/icml/ChenHH023} proposes to leverage graph sparsity during the denoising diffusion process to generate large-scale networks. 
It only focuses on a small subgraph and modifies edges between its nodes, making it much more efficient than other competing methods.
There are also efforts focused on the generation of networks with unique properties, which require specific model designs. For example, the generation of periodic graphs, such as crystal nets and polygon mesh, holds significant potential for real-world applications in areas such as material design and graphics synthesis. Wang et al.~\cite{wang2022deep} leverage variational autoencoder (VAE) to disentangle local and global patterns of networks. They design a new objective to guarantee network periodicity by ensuring the invariance of the local pattern representations of networks that contain the same local structure.

\subsection{Decision-making and control problem} \label{app:m6}
\textbf{Network control.}
In this survey, we cover three typical network control scenarios, including epidemic control, influence maximization, and coupled oscillator synchronization.
Influence maximization is a typical task in network control, which aims to determine a subset of nodes, often referred to as seed nodes, that, when activated or influenced, will lead to the largest possible diffusion of influence throughout the network.
Influence can take various forms, such as the adoption of an innovation, the spread of information, or the endorsement of an idea.
The process typically involves modeling the network topology and the dynamics of influence propagation. 
Various mathematical models, such as the Independent Cascade Model or the Linear Threshold Model, are commonly used to simulate how influence spreads through the network based on node activations~\cite{meirom2021controlling}. 
In practice, influence maximization has applications in viral marketing, social network analysis, and the design of strategies for information diffusion.

Various machine learning approaches~\cite{schuetz2022combinatorial,wang2022towards,ireland2022lense,manchanda2019learning} were proposed to solve combinatorial optimization problems, which can be employed for the influence maximization task.
The main challenge in influence maximization lies in the demand for robust generalization ability, as recalculating seed node sets for a new network is a time-consuming process.
Chen \textit{et al}~\cite{chen2021contingency} proposed an RL model with state-abstraction and reward shaping, to faciliate the transfer from training networks to unseen networks test phase.
Most works leave the information propagation process to the environment and the agent only perceives the process through reward values, which leads to insufficient modeling of the underlying network dynamics.
Ling \textit{et al}~\cite{ling2023deep} proposed DeepIM with two learning-based diffusion models to capture the information spread over the network, effectively addressing the challenge of dynamics modeling.

\section{Application details}
\subsection{Ecology network} \label{app:a1}
\subsubsection{Datasets.}
We categorize ecological networks into three distinct types mentioned above, including food webs, mutualistic networks, and comprehensive ecosystem networks. For each category, we detail their respective sources, sizes, and descriptions in Table~\ref{tab:eco}.

\subsubsection{Applications.}
Here we classify applications concerned with ecological network researchers into prediction, inference, simulation, and decision-making \& control tasks, according to task categories in Section~\ref{sec:pam}, which are shown in Table~\ref{tab:ecotask}, providing possible AI-based research methods to address these challenges.

\textbf{Prediction.}
Prediction is one of the most important problems in ecological networks. Existing efforts have a variety of prediction targets. Saavedra et al.~\cite{saavedra2013estimating} develop a theoretical framework to predict the tolerance levels of species in mutualistic networks by changing the strength of their interspecies interactions. It helps to identify scenarios in which generalist species might be the least tolerant to these changes. 
Ratzke et al.~\cite{ratzke2020strength} predicts the strength of interaction changes between bacterial species under different concentrations of available nutrients. They find that high nutrient levels lead to more intense chemical alterations in the environment, fostering predominantly negative interactions, and the stronger interactions result in a decrease in biodiversity and a reduction of community stability.
Perkins et al.~\cite{perkins2022consistent} investigate the ratio of predator-to-prey biomass across various ecosystems, analyzing 141 food webs from freshwater, marine, and terrestrial environments. It identifies a consistent sub-linear scaling pattern where predator biomass increases with prey biomass at a near $\frac{3}{4}$-power exponent, indicating that more prey biomass supports proportionally less predator biomass. Their findings provide insights into the fundamental dynamics governing energy transfer in ecological communities.
Zhao et al.~\cite{zhao2023relationships} examine the joint effects of temperature and biodiversity changes on the ecological stability of planktonic food webs, where the stability is assessed in terms of both structural and temporal aspects. The authors find that warmer temperatures are associated with stable-reduced ecological networks.

\textbf{Inference.}
Inference problems with ecological networks are usually related to their topologies, that is, inferring the original ecological network topologies or their changes under shifting environmental factors. 
Young et al.~\cite{young2021reconstruction} introduce a Bayesian statistical technique to infer network structure such as plant-pollinator network along with ecological metrics from noisy and error-prone observational data, which enables more rigorous statistical analyses of ecological variables and outcomes.

\textbf{Simulation}
Based on ecological networks, simulation tasks are mainly concerned with simulating the dynamic changes of species within ecosystems.
Bischof et al.~\cite{bischof2020estimating} utilize collected DNA samples from brown bears, gray wolves, and wolverines to track and forecast wildlife population dynamics, which provides quantitative evidence of large carnivore recovery of the system and identifies humans as the primary factor influencing the dynamics of these apex predators.
Rangel et al.~\cite{rangel2018modeling} develop a complex mechanistic simulation model to explore the drivers of biodiversity patterns in South Africa.
Pacheco et al.~\cite{pacheco2019costless} investigate the role of costless metabolite exchange in fostering stable cooperation within microbial communities using genome-scale metabolic models. The authors conducted over two million growth simulations involving 24 species in various environments and identified a broad range of metabolites that can be secreted without fitness cost, which leads to a prevalence of stable ecological network motifs. 

\textbf{Control}
One core problem of decision-making and control in ecological networks is about enhancing their resilience 
or reviving, which aims to enhance the capability of networks to persist in functionality under various perturbations.
For example, Jiang et al.~\cite{jiang2019harnessing} focus on complex ecological networks, particularly mutualistic networks between pollinators and plants, that can exhibit tipping points leading to global extinction events. The authors propose a strategy to manage and control these tipping points by maintaining a constant abundance of a specific pollinator species, effectively eliminating the hysteresis associated with tipping points preventing sudden and drastic changes in the ecosystem.

\subsection{Biology network} \label{app:a2}
\subsubsection{Datasets.}
We categorize biology networks into the above two meta-types and demonstrate their sources, sizes, and descriptions in Table~\ref{tab:bio}.

\subsubsection{Applications.}
According to AI task categories in Section~\ref{sec:pam}, we classify applications concerned by biology network researchers into 
representation, prediction, inference, and generation tasks, which are shown in Table~\ref{tab:biotask}

\textbf{Representation.} 
Representation learning for proteins is a prerequisite for understanding protein functions, interactions, and roles in health and disease.  In this field, 3D Convolutional Neural Networks (3DCNNs) and sequence-based methods and have been applied to extract representations of protein structures~\cite{ragoza2017protein, sato2019protein, pu2019deepdrug3d}. Moreover, surface-based geometric deep learning methods, which learn the patterns of chemical and geometric features that depict interaction modes of proteins, are also proven to be effective. For molecule networks, nodes represent atoms while edges are chemical bonds between them. Junction trees are widely-accepted structure representations, in which nodes correspond to molecular substructures, offering a more granular view of the molecular structure. Therefore, GNN emerges as a particularly apt choice, which can effectively capture both local and global chemical properties of molecules. For RNAs, strings are usually used to represent their secondary structures. Recurrent Neural Networks (RNNs) and Transformer-based language models~\cite{vaswani2017attention, DBLP:conf/naacl/DevlinCLT19} are promising ways to learn their representations for downstream tasks, which are capable of processing sequential data and capturing dependencies in RNA sequences.

\textbf{Prediction.}
Prediction problems with biological networks are often related to the outcomes of gene expression and cell properties, which enable the unraveling of regulatory mechanisms, identification of therapeutic targets, and the development of personalized medicine approaches. Roohani et al.~\cite{roohani2023predicting} utilize a gene interaction knowledge graph to predict transcriptional responses to both single and multigene perturbations using single-cell RNA-sequencing data, achieving superior precision in predicting genetic interactions and facilitating the design of perturbational experiments. Kaimimoto et al.~\cite{kamimoto2023dissecting} employs computational perturbations to predict shifts in cell identity with multimodal data from single-cell omic, allowing for the systematic and intuitive interpretation of transcription factors in regulating cell identity. It has been applied to various biological systems, such as haematopoiesis in mice and humans, and zebrafish embryogenesis, demonstrating its ability to infer and interpret cell-type-specific GRN configurations. Other researchers concentrate on predicting protein interactions. 
Wang et al.~\cite{wang2023deep} introduce UniBind to understand how viral variants impact protein-protein binding. It represents proteins as graphs at the residue and atom levels, integrating protein three-dimensional structure and binding affinity.

\textbf{Inference.}
For inference problems among biology networks, some studies pay attention to infer transcription factor regulatory networks (TRNs), and gene regulatory networks (GRNs). Li et al.~\cite{li2022inferring} propose to leverage GNNs for inferring TRNs from single-cell chromatin accessibility data, revealing key insights into tissue development and tumorigenesis with high efficacy even in datasets with limited cell numbers. Shu et al.~\cite{shu2021modeling} propose a generative structural equation model (SEM) to infer GRNs and generate biologically meaningful representations from single-cell RNA sequencing data. Ma et al.~\cite{ma2023single} infer GRNs and gene association networks simultaneously for each cell type from single-cell multi-omics data, outperforming existing tools in cell clustering and network construction.

\textbf{Generation.}
Generation tasks in biology networks are also a widely studied topic. Existing works mainly focus on generating 3D structures of molecules from 2D networks, which is an essential problem in the field of computational chemistry. Shi et al.~\cite{shi2021learning} propose a score-based generative model to generate stable 3D molecular conformation structures by directly estimating the gradient fields of the log density of atomic coordinates, which facilitates the generation of stable conformations through Langevin dynamics.
Luo et al.~\cite{luo2022autoregressive} propose an autoregressive flow model that creates 3D molecular structures in a step-by-step fashion, incorporating spherical message passing and an attention mechanism for extracting conditional information. Instead of directly generating 3D coordinates, G-SphereNet generates the positions of atoms by calculating distances, angles, and torsion angles, ensuring both invariance and equivariance, which has shown superior performance in tasks of random molecular geometry generation and targeted molecule discovery.

\subsection{Urban network} \label{app:a3}
\subsubsection{Datasets}

As previously mentioned, delving into urban networks and understanding their influence is instrumental for predicting the effects of disasters or accidents, revealing urban vulnerabilities, etc. Consequently, numerous public datasets on urban networks have been published. Specifically, these datasets cover topologies and dynamics of electric networks, mobile networks, road networks, and other urban transportation networks (e.g., train, airport), which have facilitated the investigation of prediction, generation, and control problems within these urban infrastructure networks.
On the other hand, a number of datasets of urban activity networks have been released to promote the study of critical research problems concerning human behavior and infectious diseases.
Overall, we have summarized these datasets in Table~\ref{tab:dataset_unet}.

\subsubsection{Applications} For urban networks, we also classify applications, ML problems, and AI technology of related works in Table~\ref{tab:urbantask}. Specifically, ML problems in these studies mainly involve prediction, generation, and control, which will be introduced in detail in the following.

\textbf{Prediction:} In urban networks, one of the most crucial applications of AI algorithms is prediction.
More specifically, prediction tasks involve cascading failures prediction, travel time estimation, and demand prediction of electric networks~\cite{varbella2023geometric,bassamzadeh2017multiscale}. AI techniques include GCN~\cite{varbella2023geometric} and Bayesian Networks~\cite{bassamzadeh2017multiscale} are leveraged to predict the future state of the whole network after a long-term evolution, or the missing information~(edge-level) of the network. 

\textbf{Simulation:} It is practically valuable to simulate traffic flow of road networks~\cite{chen2020autoreservoir} and contagion dynamics based on human contact networks~\cite{murphy2021deep,fu2023privacy}. Among these applications, the widely used AI techniques include GCN~\cite{murphy2021deep,fu2023privacy}, neural ODE~\cite{huang2021coupled}, and auto-reservoir Neural Networks~\cite{chen2020autoreservoir}. Most of them focus on simulating the future state of different urban infrastructure networks or urban activity networks based on their current states and topology. This problem is solved by proficiently modeling the dynamics within urban networks, and GCN has been validated as a robust tool for handling this problem. Specifically, Murphy et al. \cite{murphy2021deep} show the contagion dynamics on complex networks can be efficiently modeled by GCN.  
Meanwhile, Neural ODE, as a method capable of effectively modeling dynamics by using powerful neural networks to fit the derivatives of states over time, has become a potent approach for solving the prediction problem on urban networks combined with GCN methods~\cite{huang2021coupled}.

\textbf{Generation:} The generation task in this domain mainly focuses on synthesizing plausible urban networks to support downstream applications, e.g., urban infrastructure design~\cite{kocayusufoglu2022flowgen}, etc.
Specifically, 
Kocayusufoglu et al.~\cite{kocayusufoglu2022flowgen} focus on generating realistic flow graphs of urban infrastructure networks, and they propose a two-step approach based on Generative Adversarial Networks (GAN)  to generate both the adjacency matrix and flow matrix of urban infrastructure simultaneously.
Simini et al.~\cite{simini2021deep} extend the classical model for mobility flow based on feed-forward neural networks to generate plausible urban mobility networks. 
Rong et al.~\cite{rong2023complexity} further tackle this city-wide flow generation problem through the network perspective, proposing a graph denoising diffusion model that can directly generate a realistic urban mobility network.

\textbf{Decision-making and control:} 
The control problem in urban networks aims to address how to intervene in the nodes or edges of the network to influence the behavior of the entire network or reveal its key characteristics such as resilience and vulnerable nodes~\cite{mao2023detecting,hao2022reinforcement,hao2023gat}. 
Specifically, Mao et al.~\cite{mao2023detecting} utilize reinforcement learning combined with GCN to capture the risk of cascade failure and discover vulnerable infrastructures of cities. Hao et al.~\cite{hao2022reinforcement,hao2023gat} propose a reinforcement learning method combined with graph neural networks for efficient COVID-19 vaccine allocation in large-scale metropolises, addressing challenges in decision-making complexity and heterogeneous information utilization. Specifically, in the GAT-MF model proposed in~\cite{hao2023gat}, they model the interactions between agents as interactions between agents and a weighted mean field, which is based on the classical mean-field theory, to reduce the computational complexity significantly. These studies stand as a noteworthy case of applying AI technology to resolve control problems in complex networks.

\subsection{Society network} \label{app:a4}
\subsubsection{Dataset}

Based on the domains, we divide the datasets regarding society networks into three categories, including social networks, scientific collaboration networks, and economic networks. We summarize them in Table~\ref{tab:soc}. For the social network, we have three typical datasets. 
Specifically, the Twitter dataset~\cite{bovet2019influence} contains over 4 million users and 76 million (re)tweets, where social networks are extracted from their retweet behaviors. With this dataset, we can simulate political opinion dynamics and capture polarization patterns in real-world scenarios. Considering the dataset has been made publicly available recently, its AI applications remain largely underexplored. 
The Weibo datasets~\cite{xie2021detecting}, involving 192,749 Weibo users and track of 253 information items, offer both network topology and the information spreading flows over edges. This makes it a valuable dataset for information propagation tasks. Other social network datasets include Beidian~\cite{xu2019relation}, which can facilitate social recommendations. FakeNewsNet~\cite {shu2020fakenewsnet} which is collected from fake news detection.

On the other hand, compared with social networks, there are few open datasets in the domain of economic and scientific collaboration networks. For example, most supply chain datasets and the Web of Science dataset are not open to the public~\cite{xu2019resiliency,zhang2022estimating,web-of-science}. It is worth noting that Microsoft Academic Graph~\cite{mag} and OpenAlex~\cite{priem2022openalex} are two valuable datasets, which can facilitate various tasks, e.g., linking prediction~\cite{zhang2019oag}.

\subsubsection{Applications} According to the classification of AI tasks, we classify applications into three categories, including prediction, simulation, and decision-making \& control.

\textbf{Prediction.} Prediction is one of the most important applications in complex society networks. Current studies have designed various prediction targets, e.g., fake news and accounts~\cite{wu2023decor}, information propagation~\cite{yang2021full}, critical link prediction~\cite{zhang2019oag,li2023learning} and friendship prediction~\cite{zhang2023attentional}, etc. Specifically, the task of information propagation aims to predict the number of users who have known an information item from the macroscopic perspective or to predict the next user who will know the item in the microscopic view~\cite{yang2021full}. Therefore, how to utilize multi-scale information, at both macroscopic and microscopic levels, is the core of this long-standing problem in the current stage. For example, Yang et al.~\cite{yang2021full} propose a full-scale diffusion prediction model with RL modeling the macroscopic diffusion size information and RNN capturing microscopic diffusion target. Ji et al.~\cite{ji2023community} develop a community-based graph learning method, which explicitly captures information propagation. Friendship prediction, also known as social recommendations, is to predict whether two users share a social relation between them~\cite{zhang2023attentional}. Currently, Zhang et al.~\cite{zhang2023attentional} also attend to co-evolving community dynamics and propose an attention-based GNN model to achieve better performance in the task of friendship prediction. Researchers also mine social network structures to boost tasks that are traditionally believed to be unrelated to networks~\cite{wu2023decor}. For example, Wu et al.~\cite{wu2023decor} consider the degrees of news article nodes and propose a graph-learning method for fake news detection. Besides social networks, researchers also frame various prediction tasks on scientific collaboration networks~\cite{zhang2019oag} and supply chain networks~\cite{li2023learning}. For exmample, Zhang et al.~\cite{zhang2019oag} predict the links between entities from different sources in the Microsoft Academic Graph. Li et al.~\cite{li2023learning} utilize spatial-temporal knowledge graphs of the supply chain to predict links among Small and Medium-sized Enterprises.

Overall, prediction is the most fundamental task in complex society networks. Recent studies on prediction mainly focus on (i) how to utilize multi-scale information~\cite{yang2021full,ji2023community,zhang2023attentional} and (ii) how to model evolving relationship~\cite{zhang2023attentional,li2023learning}. Further work should consider how to model both of them in a unified framework.

\textbf{Simulation.} Simulation on complex society networks has become an emerging field in recent years. Prior studies mainly adopt mechanistic models as the basis and perform agent-based simulations on them~\cite{liu2023emergence}. Currently, researchers adopt a data-driven approach to better capture social dynamics~\cite{okawa2022predicting}. For example, Okawa et al.~\cite{okawa2022predicting} leverage the power of sociologically-informed neural networks and neural ordinary differential equations for modeling opinion dynamics. On the other hand, the recent development of large language models (LLM) has also largely empowered the simulation on social networks~\cite{park2023generative,gao2023s}. By encoding social mechanisms in natural languages instead of parameters and functions, these LLM-based simulations can reproduce human social behaviors in a more fine-grained manner~\cite{park2023generative,gao2023s}. 

To sum up, the recent development of AI empowers simulation methods with (i) the capability to learn real-world dynamics from data and (ii) the capability to make agents more similar to human beings. Future work should consider combining these two directions together, that is, adopting real-world dynamics to generate reliable dynamical trajectories and using LLM to enrich these trajectories.

\textbf{Decision-making and control.} Decision-making and controlling is also one ultimate target for studying complex society networks~\cite{fan2020finding,chen2021contingency,ling2023deep}. Recent work mainly focuses on two tasks: (i) Network dismantling, removing certain nodes or edges to destroy social functionalities of complex networks~\cite{fan2020finding} and  (ii) influence maximization, selecting certain nodes or edges as the seed and maximizing the final information~\cite{chen2021contingency,ling2023deep}. For example, Fan et al.~\cite{fan2020finding} introduce an RL-based framework for finding key nodes in the 9/11 terrorist networks. Chen et al.~\cite{chen2021contingency} also leverage the power of RL to find a subset of seed nodes to maximize the information spread. Focusing on the influence maximization task, Ling et al.~\cite{ling2023deep} point out the difficulties of diffusion pattern modeling and objective function solving. They propose a framework with autoencoders for modeling seed node sets, a diffusion model for capturing information spread patterns, and an objective function for handling node-centrality-constrained scenarios~\cite{ling2023deep}. Due to the heterogeneity of populations and the complexity of human behaviors, decision-making and controlling in complex society networks is a non-trivial task. Though few studies consider these aspects~\cite{ling2023deep}, future work should modeld them in a more fine-grained manner.

\end{document}